\documentclass[nohyper]{JHEP3}

\usepackage{cite}
\usepackage{epsfig}
\usepackage{amsmath}

\def\beq{\begin{equation}}
\def\eeq{\end{equation}}
\def\beqa{\begin{eqnarray}}
\def\eeqa{\end{eqnarray}}
\def\eq#1{Eq.~(\ref{#1})}

\def\M{{\cal M}}
\def\f{{\rm f}}
\def \as {\relax\ifmmode\alpha_s\else{$\alpha_s${ }}\fi}
\def\e{\epsilon}

\newcommand{\secn}[1]{Section~\ref{#1}}

\newcommand{\NP}[1]{Nucl.\ Phys.\ {\bf #1}}
\newcommand{\PL}[1]{Phys.\ Lett.\ {\bf #1}}

\newcommand{\PR}[1]{Phys.\ Rev.\ {\bf #1}}

\psfull

\title{Factorization constraints for soft anomalous dimensions in QCD scattering amplitudes}

\author{Einan Gardi \\
School of Physics, The University of Edinburgh\\
Edinburgh EH9 3JZ, Scotland, UK}
\author{Lorenzo Magnea \\
Dipartimento di Fisica Teorica, Universit{\`a} di Torino, and\\
INFN, Sezione di Torino, Via P. Giuria 1, I-10125 Torino, Italy}

\abstract{
We study the factorization of soft and collinear singularities in dimensionally-
regularized fixed-angle scattering amplitudes in massless gauge theories. 
Our factorization is based on replacing the hard massless partons by light-like 
Wilson lines, and defining gauge-invariant jet and soft functions in dimensional regularization. 
In this scheme the factorized amplitude admits a powerful symmetry: it is invariant under rescaling of individual Wilson-line velocities. This symmetry is broken by cusp singularities in both the soft and the eikonal jet functions. We show that the cancellation of these cusp anomalies in any multi-leg amplitude
imposes all-order constraints on the kinematic dependence of the corresponding soft anomalous dimension, relating it to the cusp anomalous dimension. 
For amplitudes with two or three hard partons the solution is unique: the constraints fully determine 
the kinematic dependence of the soft function. For amplitudes with four or more hard partons we present a minimal solution where the soft anomalous dimension is a sum over colour dipoles, multiplied by the cusp anomalous dimension. In this case additional contributions to the soft anomalous dimension at three loops or beyond are not excluded, but they are constrained to be functions of conformal cross ratios of kinematic variables.
}


\preprint{
Edinburgh/2008/35\\
DFTT--29/2008
}

\begin{document}

\section{Introduction}
\label{intro}

Studies of infrared and collinear singularities of fixed-angle scattering 
amplitudes in massless gauge theories have a long history (for early
results, see for example~\cite{Callan:1974zy} and~\cite{Sen:1982bt}),
and they have led to remarkable insights into the all-order structure of
the perturbative expansion.

These studies are not motivated by a purely theoretical interest: in fact, a 
detailed understanding of the long-distance singularity structure of QCD 
amplitudes is a crucial element in predicting high-energy collider cross 
sections. Indeed, the calculation of observable cross sections involves 
intricate cancellations of soft and collinear singularities between real and 
virtual corrections (see for example~\cite{Catani:1996jh}). Furthermore, with 
a precise knowledge of singularities one can predict dominant higher--order 
corrections, and in many occasions resum certain classes of logarithmically
enhanced contributions to all orders~\cite{Sterman:1995fz,Sterman:2004pd}. 

Our understanding of long--distance singularities is based on the ideas of 
factorization and universality~\cite{Collins:1989gx}. Fixed-angle scattering
amplitudes are functions of Lorentz-invariant combinations of external 
momenta, which are assumed to be uniformly much larger than the relevant
infrared cutoff, typically given by the scale of confinement; one expects then 
that exchanges of virtual particles with vanishingly small energies, or with 
vanishing transverse momenta with respect to given external legs, should 
decouple from hard exchanges, which happen at much shorter distances. 
Such a decoupling is far from apparent in Feynman diagram calculations, 
but it can indeed be proven to all orders in perturbation theory, once
gauge invariance is enforced by means of appropriate Ward 
identities~\cite{Sterman:1995fz}. 

The result has a simple structure, explained in detail in \secn{facto}. 
Briefly, multi-leg amplitudes can be organized as vectors,
in a vector space spanned by the irreducible representations of the gauge 
group that can be constructed with the given external particles; such a 
vector can be shown to have a factorized structure: each external leg
is dressed by virtual collinear emissions, building up a colour-singlet `jet' 
function; soft gluons exchanged at wide angles are assigned to a separate 
factor, which is a matrix, mixing the available colour representations; this 
matrix of `soft' functions is then contracted with a vector of hard scattering 
coefficients, which contain no infrared or collinear singularities. 

Given this factorized structure, one may immediately deduce that 
the various factors in the amplitude obey simple evolution equations, 
embodying the consequences of renormalization group as well as gauge
invariance~\cite{Contopanagos:1996nh} (for reviews of this viewpoint,
see~\cite{Laenen:2004pm,Magnea:2008ga}). Evolution equations of this 
type were derived for the first time for the form factors of elementary fields, 
with a variety of methods~\cite{Mueller:1979ih,Collins:1980ih,Sen:1981sd},
and were later extended to Wilson loops~\cite{Polyakov:1980ca,Dotsenko:1979wb,Brandt:1981kf,Korchemsky:1985xj,Ivanov:1985np,Korchemsky:1987wg,Korchemsky:1988hd,Korchemsky:1988si} and to cross sections 
and amplitudes of phenomenological interest~\cite{Collins:1981uk,Korchemskaya:1994qp,Botts:1989kf,Kidonakis:1997gm}. 
Solving these equations leads to the exponentiation of all infrared and 
collinear singularities. The singularities in the exponent are generated by 
integrals over the scale of the running coupling of specific anomalous 
dimensions, which can be computed order-by-order in perturbation theory. 
A significant step was taken in~\cite{Magnea:1990zb}, where the evolution
equation for the Sudakov form factor was solved in dimensional regularization.
Within this framework, infrared and collinear poles are generated by integration 
over the scale of the $D$-dimensional version of the running coupling; the results 
of exponentiation can then be directly compared with finite-order Feynman 
diagram calculations; the Landau pole is also regulated by dimensional 
continuation, so that resummed amplitudes can be computed as analytic 
functions of the coupling at a fixed scale and of the dimension of 
space-time~\cite{Magnea:2000ss}. This approach was extended 
to multi-leg amplitudes in~\cite{Sterman:2002qn}, confirming earlier
predictions~\cite{Catani:1998bh} on the structure of singularities at NNLO. 

In recent years, the development of novel and advanced techniques for 
high-order calculations in QCD and in general gauge theories has stimulated
further investigation of the exponentiation of infrared and collinear singularities.
In particular,  great theoretical effort has been made to further our understanding
of amplitudes in supersymmetric gauge theories, and most notably in the 
maximally supersymmetric ${\cal N} = 4$ Yang-Mills theory. This theory 
is of special interest for several reasons: it is quantum conformal invariant; 
in the planar limit, it is expected to have a simple, theoretically accessible 
strong-coupling limit, because of its connection with string theory 
through the AdS-CFT correspondence~\cite{Maldacena:1997re}; finally, 
its amplitudes and anomalous dimensions are of practical relevance, since 
they have nontrivial relations with the corresponding quantities in QCD (see for 
example Refs.~\cite{Kotikov:2004er,Dixon:1996wi}, and recent studies of the Regge limit~\cite{DelDuca:2008jg,Bartels:2008sc}). Explicit calculations for the 
four-point function in ${\cal N} = 4$ super-Yang-Mills (SYM) theory have led to an 
all-order conjecture~\cite{Bern:2005iz}, suggesting that non-singular terms 
exponentiate together with infrared and collinear poles, at least for the class of 
maximally helicity violating (MHV) amplitudes. While this conjecture has now been 
shown to fail starting with the two-loop six-point function~\cite{Bern:2008ap}, it is 
clear that ${\cal N} = 4$ SYM perturbative amplitudes must have a remarkably 
simple all-order structure, which may well be brought under full theoretical 
control in the near future. A step in this direction was taken with the discovery
of a surprising duality between scattering amplitudes in momentum space 
and expectation values of Wilson loops taken in an auxiliary coordinate 
space~\cite{Korchetal}. Further, remarkable progress was made at strong 
coupling in Ref.~\cite{Alday:2007hr}, where a calculation of the four-point
amplitude  was performed, by adapting string techniques to dimensional 
regularization. This allowed a direct comparison with resummed perturbative 
calculations, finding an exact matching in the structure of long-distance 
singularities. Recent results in this fast-developing field are reviewed 
in Ref.~\cite{Alday:2008yw}.

Most of the calculations just described have been carried out in the planar 
limit\footnote{An exception is Ref.~\cite{Naculich:2008ys}, 
which studies the leading infrared singularities in subleading colour 
components of the ${\cal N}=4$ SYM gluon-gluon scattering amplitude to 
three-loop order.}, which has special simplifying properties. 
In this limit, soft contributions to multi-leg amplitudes 
can be further factorized into a product of `wedges', each one proportional 
to a form factor, since soft exchanges can only take place between adjacent 
external legs. In essence, in the planar limit amplitudes can have only a single
colour structure, so that the soft anomalous dimension matrix must be 
proportional to the unit matrix. All soft and collinear singularities are then 
determined by just two colour-diagonal functions: the cusp anomalous dimension 
$\gamma_K (\as)$~\cite{Korchemsky:1985xj,Ivanov:1985np,Korchemsky:1987wg,Korchemsky:1988hd,Korchemsky:1988si}, and a subleading function 
$G(\as)$~\cite{Dixon:2008gr}, responsible for single soft or collinear poles. 

It is of great interest to push our understanding of infrared singularities in terms 
of a limited set of anomalous dimensions beyond the planar limit. Indeed, from a 
theoretical point of view, only at non-planar level one begins to see the intricate 
pattern of colour correlations that are characteristic of non-abelian gauge 
theories: only at this level space-time and colour degrees of freedom become
explicitly correlated. Furthermore, colour-subleading contributions in QCD have 
important phenomenological effects on resummed hadronic cross sections, 
beginning at the next-to-leading logarithmic order, and the understanding of
subleading poles would also play an important role in the development of infrared and collinear subtraction schemes at higher orders in perturbation theory.
Finally, recent work~\cite{Dixon:2008gr,Eynck:2003fn,Laenen:2005uz,
Laenen:2008ux,Gardi:2007ma,Gardi:2005yi, Gardi:2006jc} 
has highlighted new properties of the functions that generate 
infrared and collinear enhancements in gauge theory amplitudes and cross 
sections in the case of two hard partons, leading to a better understanding 
of the process--dependence of soft radiation, to the discovery of all-order
connections between different physical processes, and to the possibility of
performing internal resummations of running--coupling corrections within the
Sudakov exponent. It would be very interesting to extend these studies to 
general colour configurations.  

Soft anomalous dimension matrices for multi-particle 
scattering have also been intensively studied in recent years. A complete 
one-loop calculation for the simplest non-trivial case of $2 \rightarrow 2$ 
scattering was carried out originally in~\cite{Kidonakis:1998nf}. More 
recently, the calculation was reproduced in a physically motivated, 
dipole-based formalism in~\cite{Dokshitzer:2005ig}: an interesting 
observation there was that the anomalous dimension matrix for gluon-gluon
scattering displays an unexplained symmetry relating kinematic invariants 
with the number of colours~$N_c$. A different symmetry property was 
observed by~\cite{Seymour:2005ze}, where it was noted that all one-loop 
anomalous dimension matrices are complex symmetric matrices in a suitably 
chosen orthonormal basis. This property was later explicitly verified with
the calculation of the matrices for all $2 \rightarrow 3$ processes at one 
loop~\cite{Kyrieleis:2005dt,Sjodahl:2008fz}, and very recently 
proven~\cite{Seymour:2008xr}. 

Finally, a remarkable result was derived in~\cite{MertAybat:2006mz}, 
where it was shown that soft anomalous dimension matrices at two loops, 
with any number of external legs, are proportional to their one-loop value, with 
the proportionality constant given by the two-loop coefficient of the cusp
anomalous dimension. This is of course a great reduction in the number of 
possible degrees of freedom, since a priori each matrix element could have 
acquired an independent two-loop correction. The fact that the correlation 
between colour and kinematic dependence in the soft function does not get 
more complex at two loops as compared to one loop,  calls for a deeper
explanation. At present it is not known whether this remarkable property 
remains valid at higher orders.

In this paper, we begin to tackle this question. In \secn{facto} we develop 
in detail the factorization of soft and collinear singularities for fixed--angle scattering amplitudes, following the approach of Ref.~\cite{Dixon:2008gr}. 
There are two main differences between our factorization and earlier calculations 
of soft matrices. First, we employ dimensional regularization as the unique infrared
and collinear regulator: thus, for example, in contrast with Ref.~\cite{MertAybat:2006mz} we do not tilt the Wilson lines off the light cone to regulate collinear 
poles. While this approach makes explicit loop calculations slightly more delicate, 
it has the advantage that Wilson line correlators are given by pure counterterms 
to all orders in perturbation theory, and they do not depend on any mass scales. 
Second, instead of using the jet definition as the square-root of the
Sudakov form factor as in Refs.~\cite{Sterman:2002qn,MertAybat:2006mz}, 
we define each jet $J_i$ by introducing a separate auxiliary vector $n_i$, as suggested in early work on Sudakov factorization~\cite{Collins:1989bt}. This will allow us to conveniently trace the effect of rescaling of the Wilson--line velocities.

\secn{eiko} studies the kinematic dependence of the eikonal functions
that enter the factorization of multi-parton amplitudes. First, in
\secn{eikojet}, we consider eikonal jets, and we determine their
kinematic dependence to all orders in perturbation theory in terms
of the cusp anomalous dimension. This simple result follows from the
fact that the eikonal jet is defined as a correlator of semi-infinite
Wilson lines (see (\ref{calJdef}) below) one of which goes along the
light-like direction defined by the momentum of an external hard parton.
Any such correlator of semi-infinite Wilson lines is classically
invariant under rescaling of any of the corresponding velocity vectors
(independently of whether they are light-like or not): this invariance
is a property of the eikonal Feynman rules. In the presence of cusps
with light-like rays, however, the renormalization procedure breaks
this invariance: the counter terms include double poles, corresponding
to overlapping ultraviolet and collinear singularities, along with
single poles that carry explicit dependence on the normalization of
the light-like velocity vectors. Thus, the renormalized correlators,
which do retain their invariance under rescaling of any non-light-like
Wilson-line velocity vector, acquire a dependence on the normalization
of the light-like ones. The origin of this anomaly is well understood
\cite{Polyakov:1980ca,Dotsenko:1979wb,Brandt:1981kf,Korchemsky:1985xj}:
the violation of classical rescaling invariance is governed by the
cusp anomalous dimension, and we will refer to it as the `cusp anomaly'.

In \secn{const} we extend the analysis to soft gluon 
functions. To deal with the general multi-leg case, we examine combinations 
of soft and jet correlators where the cusp anomaly cancels, so that rescaling
invariance must be recovered. We find that this strongly constrains 
the dependence of the soft anomalous dimension on the kinematics of the
scattering process, and eventually also on the colour degrees of freedom.
 
\secn{sec:two_legs} deals with the simple case of amplitudes with only two 
hard coloured partons. In \secn{conssud} we develop the consequences of the 
new constraints for the case of the Sudakov form factor, and show that the 
complete dependence of the corresponding eikonal function on the kinematics 
is indeed governed by the cusp anomalous dimension. In \secn{partojet} we
analyse the kinematic dependence of the partonic jet function and contrast it 
with the eikonal case.

In \secn{ansa} we return to the case of generic multi-parton fixed-angle 
scattering amplitudes and study the impact of the new constraints on the 
soft function. We show that while these constraints are insufficient to fully
determine the functional dependence on the kinematic variables, they admit 
a remarkably simple solution,  where the soft anomalous dimension matrix at 
any order in perturbation theory is proportional to the one-loop result. The
solution corresponds to a sum over all colour dipoles, which correlate the 
kinematic dependence to the colour degrees of freedom, multiplied by the 
cusp anomalous dimension. This formula is consistent with the result of 
Ref.~\cite{MertAybat:2006mz} at two loops and generalizes it to all orders.
We also discuss possible sources of further corrections. We 
conclude in \secn{conclu} by summarizing our results, while two appendices
discuss concrete examples. In Appendix \ref{three_partons} we describe the 
special case of amplitudes with three hard partons, where the sum-over-dipoles
formula is the unique solution to the new constraints. Appendix 
\ref{qqbar-one-loop-example} studies $2 \to 2$ scattering of quarks at one 
loop, describing the way in which conformal cross ratios are formed through 
a sum over diagrams.

\section{Factorization of fixed--angle scattering amplitudes}
\label{facto}

We begin by describing the factorization of a general fixed--angle
massless gauge theory amplitude into soft, hard and jet 
functions. We follow the notations of Ref.~\cite{Dixon:2008gr} and
generalize the definition of the soft function given there to the case
of multileg amplitudes. The amplitude ${\cal M}$ describes the scattering 
of  $n$ hard massless gauge particles (plus any number of colour--singlet 
particles) so it is characterized by $n$ colour indices $\{\alpha_i\}$, $i = 1, 
\ldots n$, belonging to arbitrary representations of the gauge group. 
The representation content of the amplitude is collectively denoted 
by $[f]$. Such a coloured object can be decomposed into components 
by picking a basis of independent colour tensors with the same index 
structure. We denote these tensors by $\left(c_L\right)_{\{\alpha_i\}}$, 
where $L = 1, \ldots N^{[f]}$ and $N^{[f]}$ is the number of irreducible
representations of the gauge group that can be constructed with the 
given particles. We write then
\beq
\label{amp}
  \M^{[\f]}_{\{\alpha_i\}} \left(p_i/\mu, \as(\mu^2),
  \epsilon \right)  = 
  \sum_{L = 1}^{N^{[\f]}} \M^{[\f]}_{L}
  \left(p_i/\mu, \as(\mu^2), \epsilon \right)
  \, \left(c_L\right)_{\{\alpha_i\}} \, ,
\eeq
with $\mu$ being the renormalization scale and $\epsilon=2-D/2$, where $D$ 
is the dimension of space-time. General factorization arguments guarantee 
that the colour components ${\cal M}_L$ of the amplitude may be written in a 
factorized form. Following~\cite{Sen:1982bt,Kidonakis:1998nf,Sterman:2002qn,MertAybat:2006mz,Dixon:2008gr}, we write
\beqa
\label{facamp}
  \M^{[\f]}_{L} \left(p_i/\mu, \as (\mu^2),
  \epsilon \right) & = & 
  {\cal S}^{[\f]}_{L K} \left(\beta_i \cdot \beta_j, \as (\mu^2), \e 
  \right) \,  H^{[\f]}_{K} \left( \frac{2 p_i \cdot p_j}{\mu^2},
  \frac{(2 p_i \cdot n_i)^2}{n_i^2 \mu^2}, \as (\mu^2) \right)
  \nonumber \\ && \times
  \prod_{i = 1}^n \frac{{\displaystyle J_i 
  \left(\frac{(2 p_i \cdot n_i)^2}{n_i^2 \mu^2},
  \as (\mu^2), \e \right)}}{{\displaystyle {\cal J}_i 
  \left(\frac{2 (\beta_i \cdot n_i)^2}{n_i^2}, \as (\mu^2), \e \right)} \,} \,\,,
\eeqa
where the hard function $H^{[\f]}_{K}$, like the amplitude $\M^{[\f]}_{L}$ 
itself, is a vector in the colour space described above; the soft function 
${\cal S}^{[\f]}_{L K}$ is a matrix in this space, while the jet functions 
$J_i$ and  ${\cal J}_i$ do not carry any colour index. A sum over
$K$ is assumed on the \emph{r.h.s.} The soft matrix ${\cal S}$ and the 
jet functions $J$ and ${\cal J}$ contain all infrared and collinear singularities 
of the amplitude, while the hard functions $H_K$ are independent of $\epsilon$. 
Each of the functions appearing in \eq{facamp} is separately gauge invariant 
and admits an operator definition given below. These definitions also clarify 
the choice of the arguments of each function. In particular, we have exhibited 
here the fact that the eikonal functions ${\cal S}$ and ${\cal J}$ depend on 
the dimensionless four-velocities $\beta_i$ associated with external particles,
rather than the particle momenta $p_i$. The velocities are defined by scaling 
the momenta $p_i$ according to $p_i =  \beta_i Q_0/\sqrt{2}$, where the
magnitude of $Q_0$ is unimportant so long as this substitution\footnote{As 
soon as the $\beta_i$ variables are used in the hard functions $H$ or in the
partonic jet functions $J$, which do depend on physical scales, $Q_0$ needs 
to be specified. We shall avoid that, except in \secn{partojet} where we relate 
the normalization of the partonic jet to the Sudakov form factor.} is restricted to 
the eikonal functions. The fixed-angle assumption implies that all scalar 
products $\beta_i \cdot \beta_j$ ($i \neq j$) are of order $1$. 

The definitions of the soft and jet functions all involve Wilson lines, which 
we write as 
\beq
  \Phi_n (\lambda_2, \lambda_1) =
  P \exp \left[\, {\rm i} g \int_{\lambda_1}^{\lambda_2} d \lambda \,
  n \cdot A(\lambda n) \, \right]~.
\eeq
In terms of these Wilson--line operators, one may then define the `partonic jet' 
functions (for, say, an outgoing quark with momentum $p$) as
\beq
  \overline{u}(p)  \, J \left( \frac{(2p \cdot n)^2}{n^2 \mu^2}, \as(\mu^2), 
  \e \right) \, = \, \langle p \, | \, \overline{\psi} (0) \, \Phi_n (0, - \infty) \,  
  | 0 \rangle\, .
\label{Jdef}
\eeq
The function $J$ represents a transition amplitude connecting the 
vacuum and a one-particle state. The eikonal line $\Phi_n$ simulates
interactions with fast partons moving in different directions: the direction
$n^\mu$ is arbitrary, but off the light-cone (in order to avoid spurious 
collinear singularities). Since eikonal Feynman rules are invariant under 
rescalings of the eikonal vector $n^\mu$, and this invariance is not broken 
by the cusp anomaly for $n^2 \neq 0$, $J$ can depend on the vectors $p$ 
and $n$ only through the argument given in \eq{Jdef}\footnote{For later convenience factors of $2$ have been introduced into the arguments of the jet functions.}. To avoid any ambiguity with respect to unitarity phases associated 
with the first argument of $J$ we shall choose $n$ such that $p\cdot n>0$. 
Note that this can be done so long as one retains the vectors $n$ corresponding 
to different partons independent of each other.
 
The factorization formula (\ref{facamp}) also requires to introduction of the 
eikonal approximation to the partonic jet $J$, which we call the `eikonal 
jet'. It is defined by
\beq
  {\cal J} \left( \frac{2 (\beta \cdot n)^2}{n^2}, \as(\mu^2), \e 
  \right)  \, = \, \langle 0 | \, \Phi_{\beta}(\infty, 0) \, 
  \Phi_{n} (0, - \infty) \, | 0 \rangle~.
\label{calJdef}
\eeq
Both the partonic jet (\ref{Jdef}) and the eikonal jet (\ref{calJdef}) have 
infrared divergences, as well as collinear divergences associated to their 
light-like leg; thus, they display double poles order-by-order in perturbation 
theory. The double-pole singularities are however the same, since in the 
infrared region ${\cal J}$ correctly approximates $J$: singular contributions 
to the two functions differ only by hard collinear radiation. 

It is important to note that the eikonal jet ${\cal J}$ depends on the
renormalization scale only through the coupling: indeed, diagram by diagram 
${\cal J}$ is given by integrals with no dimensionful parameter. 
Such integrals vanish identically in dimensional regularization, but since this 
trivial result involves cancellations between ultraviolet and infrared singularities,
upon renormalization ${\cal J}$ becomes non-trivial: the contribution of each 
graph equals minus the corresponding ultraviolet counterterm. As a consequence, using a minimal subtraction scheme, the result for ${\cal J}$ at each order 
in $\alpha_s$ is a sum of poles in $\epsilon$, without any non-negative powers. 
These properties are not special to the jet function, but apply 
to any eikonal function not involving dimensionful parameters,  provided it is 
defined in dimensional regularization and in a minimal subtraction scheme.

The final ingredient in \eq{facamp} is the soft matrix. It is constructed by 
taking the eikonal approximation for all soft exchanges. Since soft gluons
are insensitive to the structure of hard collinear emissions, they couple 
effectively to Wilson lines in the colour representation of the corresponding 
hard external parton. Such exchanges mix the colour components of the 
amplitude, so one is led to define  
\beq
\left( c_L \right)_{\{\alpha_k\}} {\cal S}^{[f]}_{L K} \left(\beta_i 
\cdot \beta_j, \as(\mu^2), \e \right) = \sum_{\{\eta_k\}} \, \, \langle 0 |  \,
\prod_{i = 1}^n \Big[ \Phi_{\beta_i} (\infty, 0)_{\alpha_k, \eta_k} 
\Big]  \, | 0 \rangle  \, \left( c_K \right)_{\{\eta_k\}} \, ,
\label{softcorr}
\eeq
where for simplicity of notation we have defined all eikonal lines as outgoing.
Note that in our definition we keep all Wilson lines on the light-cone. As a consequence, also the soft matrix is a pure counterterm in dimensional regularization and it depends on the renormalization scale only through the coupling; furthermore, the homogeneity of the eikonal Feynman rules with 
respect to rescalings of the eikonal vectors $\beta_i$ would suggest that 
${\cal S}$ can depend on $\beta_i$ only through homogeneous ratios 
invariant under such rescalings. As described in~\cite{Dixon:2008gr}, this is 
not true: indeed, rescaling invariance is broken by the cusp anomaly, so that 
the soft matrix acquires nontrivial dependence on the scalar products $\beta_i
\cdot \beta_j$. This observation will be central to our arguments in the rest 
of the paper.

The soft matrix, \eq{softcorr},  displays both infrared and collinear
poles. One must then correct the factorization formula in order to avoid 
double counting of the infrared-collinear region for each external leg. This is achieved in \eq{facamp} by dividing by an eikonal jet ${\cal J}_i$ for each 
external leg, thus removing from $J_i$ its eikonal part, which is already 
accounted for in ${\cal S}^{[f]}_{L K}$. One may then observe that the 
ratios $J_i/{\cal J}_i$ are free of infrared poles, and thus contain 
only single collinear poles at each order in perturbation theory. Similarly, the
`reduced' soft matrix
\beq
  \overline{{\cal S}}^{[f]}_{L K} \left(\rho_{i j},\as(\mu^2), \e \right) = 
  \frac{{\cal S}^{[f]}_{L K} \left(\beta_i 
  \cdot \beta_j, \as(\mu^2), \e \right)}{\displaystyle \prod_{i = 1}^n 
  {\cal J}_i \left(\frac{2(\beta_i \cdot 
  n_i)^2}{n_i^2}, \as(\mu^2), \e \right)}
\label{reduS}
\eeq
is free of collinear poles, and contains only infrared singularities originating 
from soft gluon radiation at large angles with respect to all external legs.
This means that the effects of the cusp anomaly, which is the source of 
double infrared-collinear poles, must cancel in ${\overline{\cal S}}$. More 
generally, invariance under rescaling of each individual light-like eikonal 
velocity, 
\beq
\label{rescaling}
\beta_i \to  \kappa_i \beta_i \, ,
\eeq
which is broken separately in ${\cal S}$ and in ${\cal J}$, must be 
recovered in their ratio, \eq{reduS}. Indeed, in the factorized amplitude 
(\ref{facamp}) the dependence on the normalizations of the vectors 
$\beta_i$  appears only though the eikonal functions contained in 
\eq{reduS}, so the invariance of the amplitude as a whole with respect 
to such rescaling amounts to invariance of ${\overline{\cal S}}$. 
The immediate consequence is that $\overline{{\cal S}}$ can only 
depend on arguments that are simultaneously homogeneus in 
$\beta_i$ and in $n_i$. Given the different functional dependencies 
of ${\cal S}$ and ${\cal J}$, this can be achieved only if $\overline{{\cal 
S}}$ depends on kinematics only through the variables\footnote{Following 
\cite{Catani:1998bh} we keep track of the unitarity phase by writing $-\beta_i
\cdot \beta_j = \left|\beta_i\cdot\beta_j\right|\,{\rm e}^{{\rm i}\pi 
\lambda_{ij}}$ where $\lambda_{ij}=1$ if $i$ and $j$ are both initial-state 
partons or are both final--state partons, and $\lambda_{ij}=0$ otherwise.}
\beq
\rho_{ij} \equiv \frac{ n_i^2 \, n_j^2 \, \, (\beta_i \cdot \beta_j)^2 \,
{\rm e}^{2\pi{\rm i}\,\lambda_{ij}}}
{4 \,(\beta_i \cdot n_i)^2 (\beta_j \cdot n_j)^2} \,  .
\label{rhoij}
\eeq
In \secn{const} we will explore further consequences of this constraint on 
the functional dependence of the reduced soft matrix.

Finally, it is important to control the ultraviolet behavior of the jet and soft
functions thus introduced. All these functions are multiplicatively renormalizable~\cite{Dotsenko:1979wb,Brandt:1981kf};
there is however an important difference between eikonal operators involving 
light-like Wilson lines and partonic amplitudes. First, as already mentioned, the
ultraviolet divergence of eikonal operators is directly related to their infrared
singularities. Moreover, anomalous dimensions of operators involving cusped 
Wilson lines with light-like segments are themselves divergent, due to the
overlapping of collinear and ultraviolet poles. These divergences are controlled 
by the the cusp anomalous dimension~\cite{Korchemsky:1985xj,Ivanov:1985np,Korchemsky:1987wg,Korchemsky:1988hd,Korchemsky:1988si}. 
Let us then write down renormalization group equations for the various 
functions defined above.

The partonic jet $J$ does not involve any light-like Wilson line, and therefore 
does not have a cusp anomaly. Its anomalous dimension is finite, and one 
may write
\beq
  \mu \frac{d}{d \mu} 
  \ln J_i \left( \frac{(2 p \cdot n)^2}{n^2 \mu^2}, \as(\mu^2), \e \right) 
  = - \, \gamma_{J_i} (\as(\mu^2)) \,.
\label{renJ}
\eeq
In contrast, for the eikonal jet ${\cal J}_i$ we write
\beq
  \mu \frac{d}{d \mu} 
  \ln {\cal J}_i \left( \frac{2 (\beta \cdot n)^2}{n^2}, \as(\mu^2), 
  \e \right) = - \, \gamma_{{\cal J}_i} \left(\frac{2 (\beta 
  \cdot n)^2}{n^2}, \as(\mu^2), \e \right) \, .
\label{rencalJ}
\eeq
In both cases the index $i$ is kept as a reminder that the jet function $J$ for 
a given parton $i$ carries information not only on the kinematics, but also on 
the parton spin, flavor and colour, while the eikonal jet ${\cal J}$ depends on 
the colour representation only. 

For the soft matrix ${\cal S}$ multiplicative renormalizability must be 
understood in the sense of matrix multiplication~\cite{Brandt:1981kf}, and one writes
\beq
\mu  \frac{d}{d \mu} {\cal S}_{I K}^{[f]} \left( \beta_i \cdot \beta_j, \as(\mu^2), \e \right) = - \, 
\Gamma^{{\cal S}}_{I J} \left( \beta_i \cdot \beta_j, \as(\mu^2), \e 
\right) \, {\cal S}_{J K}^{[f]} \left( \beta_i \cdot \beta_j, \as(\mu^2), 
\e \right) \, .
\label{renS}
\eeq
where $\Gamma^{{\cal S}}_{I J}$ will be referred to as the `soft anomalous dimension'; it is similar to the `cross anomalous dimension' of Ref.~\cite{Brandt:1981kf,Korchemskaya:1994qp}, taken in the limit where all the lines are light-like.   
In the following sections we shall examine the dependence of the anomalous dimensions defined above on the kinematic variables as well as on the colour degrees of freedom.

\section{On the kinematic dependence of eikonal functions}
\label{eiko}

We now discuss the properties of the eikonal functions ${\cal J}$ and
${\cal S}$ taking into account their gauge invariance, their
renormalization group evolution and their independence of any
dimensionful kinematic scale. By considering the effect of velocity
rescaling we deduce that the kinematic dependence of these functions is
tightly connected with cusp singularities. We first illustrate this for
the eikonal jet function ${\cal J}$, and then we move on to consider the
central object of our work, the soft anomalous dimension matrix
introduced in \eq{softcorr}.

\subsection{Explicit solution for the eikonal jet}
\label{eikojet}

Let us consider first the anomalous dimension of the eikonal jet in \eq{rencalJ}. 
One observes that the homogeneity of eikonal Feynman rules under the 
rescaling in \eq{rescaling} would forbid any dependence on $w_i \equiv 
2 (\beta_i \cdot n_i)^2/n_i^2$, were it not for the cusp singularity. 
One expects then that the full $w_i$ dependence of $\gamma_{{\cal J}_i}$
should be proportional to the cusp anomalous dimension, and this is indeed 
the case as we now explicitly show.

Our starting point is \eq{rencalJ}; in dimensional regularization, the statement 
that ${\cal J}_i$ is a pure counterterm implies that it can depend on $\mu$ 
only through the running coupling; one may then solve \eq{rencalJ} as
\beq
{\cal J}_i \left( w_i, \as(\mu^2), \e \right) = \exp \left[ -\frac{1}{2}
\int_0^{\mu^2} \frac{d \xi^2}{\xi^2} \, \gamma_{{\cal J}_i} \left(w_i, 
\as \left(\xi^2, \e \right), \e \right) \right] \, .
\label{solrencalJ}
\eeq
Next, we observe that the eikonal jet must obey an evolution equation of
the same form as the Sudakov form factor itself, a so-called `K+G'
equation; a similar observation was made in Ref.~\cite{Dixon:2008gr} 
concerning the eikonal approximation to the form factor. Following the 
standard reasoning, one rewrites the anomalous dimension $\gamma_{{\cal 
J}_i}$ as a sum of a singular term, generated by the cusp singularity, and a
residual finite function that contains the kinematic dependence. We write then
\beq
\label{gamma_J_split}
\gamma_{{\cal J}_i} \left(w_i, \as \left(\mu^2, \e \right), \e \right) =
- \, \frac12  \, G_{{\cal J}_i} \left(w_i, \as( \mu^2, \e) \right) \, + \, 
\frac14 \int_0^{\mu^2} \frac{d \lambda^2}{\lambda^2} \, \gamma_K^{(i)}
\left( \alpha_s (\lambda^2, \epsilon) \right) \, .
\eeq
Here we have introduced the cusp anomalous dimension $\gamma_K^{(i)} 
(\as)$, for an eikonal line in the representation of parton $i$, and a remainder
function $G_{{\cal J}_i} (w_i, \as)$. The normalization of the singular term 
on the \emph{r.h.s.} of \eq{gamma_J_split} is one half of
the corresponding term in the Sudakov form factor, since the form factor is
comprised of two jets\footnote{Indeed an alternative definition of the partonic 
jet function, which was used for example in  Refs.~\cite{Sterman:2002qn,MertAybat:2006mz}, 
is based on taking the square root of the Sudakov form factor.}.  

Upon inserting \eq{gamma_J_split} into \eq{solrencalJ}, and changing the 
order of integration, one readily arrives at
\beq
\label{calJ_after_split}
{\cal J}_i \left( w_i, \as(\mu^2), \e \right) = \exp \left[ \frac{1}{2}
\int_0^{\mu^2} \frac{d \xi^2}{\xi^2} \, \left( \frac12 
G_{{\cal J}_i} \left(w_i, \as(\xi^2, \e) \right) \, - \, \frac14 \, 
\gamma_K^{(i)} \left(\as \left(\xi^2, \e \right) \right) \,
\ln \frac{\mu^2}{\xi^2} \right) \right] \, ,
\eeq
which is analogous to the expression for the Sudakov form factor, as given in 
\eq{Sud_FF_G_plus_K} below  (or in Eq. (2.11) in Ref.~\cite{Dixon:2008gr}),
with the physical scale of the form factor, $-Q^2$, replaced here by the 
renormalization point $\mu^2$. The finite function $G_{{\cal J}_i}$ has 
no explicit $\epsilon$ dependence in a minimal subtraction scheme, since 
${\cal J}_i$ is a pure counterterm.

We are now going to show that \eq{calJ_after_split} can be further 
simplified, since the dependence on the kinematic variable $w_i$ in the 
function $G_{{\cal J}_i}$ can be completely solved for. In order to do that,
we use the results of Ref.~\cite{Dixon:2008gr} for the $w_i$ dependence of 
the eikonal jet, which is given by
\beq
w_i \frac{\partial}{\partial w_i} \ln {\cal J}_i \left( w_i, \as(\mu^2 ), 
\e \right)
\, = \, - \frac{1}{8} \int_0^{\mu^2} \frac{d \xi^2}{\xi^2} \, 
\gamma_K^{(i)} \left(\as  \left(\xi^2, \e \right) \right) \, .
\label{dcalJdu}
\eeq
Clearly, \eq{calJ_after_split} and \eq{dcalJdu} are compatible only if
$G_{{\cal J}_i} \left(w_i, \as \right)$ is a linear function of $\ln w_i$. 
Indeed,  by taking the derivative of \eq{calJ_after_split} with respect to 
$\ln w_i$, and using \eq{dcalJdu}, one gets
\beq
w_i \frac{\partial}{\partial w_i} \int_0^{\mu^2} \frac{d\xi^2}{\xi^2}
G_{{\cal J}_i}\left(w_i,\as(\xi^2,\e)\right) = -\frac{1}{2} \int_0^{\mu^2} 
\frac{d \xi^2}{\xi^2} \, \gamma_K^{(i)} \left(\as  \left(\xi^2, \e \right) 
\right) \, ,
\label{intcond}
\eeq 
for any $\mu^2$. Therefore
\beq
w_i \frac{\partial}{\partial w_i} G_{{\cal J}_i}\left(w_i,\as(\xi^2,\e)
\right) = -\frac{1}{2} \, \gamma_K^{(i)} \left(\as  \left(\xi^2, \e \right) 
\right) \, ,
\label{diffcond}
\eeq 
which we integrate to get
\beq
\label{G_calJ}
G_{{\cal J}_i} \left(w_i , \as \right) = - \frac{1}{2} \, 
\gamma_K^{(i)} \left( \as \right) \, \ln(w_i) \,
+ \, \delta_{{\cal J}_i} ( \as ) \, ,
\eeq
where $\delta_{{\cal J}_i}$ is a constant of integration, free of any 
kinematic dependence. Using \eq{G_calJ} in \eq{gamma_J_split} we finally 
get
\beqa
\gamma_{{\cal J}_i} \left( w_i, \as(\mu^2,\e), \e \right) & = &
- \frac12 \, \delta_{{\cal J}_i} \left(\as(\mu^2,\e) \right) \, 
+ \frac14 \gamma_K^{(i)} \left( \alpha_s(\mu^2, \e) \right) \, \ln (w_i) \,
\nonumber \\ && + \,  
\frac14 \int_0^{\mu^2} \frac{d \xi^2}{\xi^2} \, \gamma_K^{(i)} \left(
\alpha_s (\xi^2,\epsilon) \right) \, .
\label{fingamJ}
\eeqa
We can now write down our final expression for the eikonal jet, using
\eq{calJ_after_split}. We obtain
\beq
{\cal J}_i \left(w_i, \alpha_s (\mu^2), \epsilon \right) =
\exp\Bigg\{
\frac12 \int_{0}^{\mu^2} \frac{d \lambda^2}{\lambda^2}
\bigg[ \frac12 \delta_{{\cal J}_i} \Big( \as( \lambda^2, \epsilon ) \Big)
- \frac14 \gamma_K^{(i)} \Big( \as(\lambda^2, \epsilon)
\Big) \, \ln \left(\frac{w_i \mu^2}{\lambda^2} \right)
\bigg] \Bigg\} \, ,
\label{fincalJ}
\eeq
where, as anticipated, the kinematic dependence of the eikonal jet is explicitly written, to all orders in perturbation theory, in terms of the cusp anomalous dimension. We observe that the cusp anomalous dimension simultaneously 
controls the double poles and the kinematic dependence of the single poles. 
In the following sections we will see that this property holds also in the more
complex soft functions. 

Returning to the comparison with the Sudakov form factor 
case (see \eq{Sud_FF_G_plus_K}), we now see that the physical scale $-Q^2$ 
is replaced here by $\mu^2 w_i = 2 \mu^2 (\beta_i\cdot n_i)^2/n_i^2$. 
It is important to note that \eq{fincalJ} can also be expressed as
\beq
{\cal J}_i(w_i,\alpha_s(\mu^2),\e) =
\exp\left\{\frac12 \int_0^1 \frac{d\theta}{\theta} \left[
\frac12 \delta_{{\cal J}_i}(\alpha_s(\mu^2\theta,\e)) 
-\frac14 \gamma_K^{(i)}(\alpha_s(\mu^2\theta,\e))\ln
\left(\frac{w_i}{\theta}\right)\right]
\right\}\,,  
\eeq
exhibiting the fact that $\mu$ dependence appears only through the
$D$-dimensional running coupling.

Finally we emphasize that the above result for the eikonal jet holds for quarks as well as for gluons. In fact, the  dependence of \eq{fincalJ} on the colour representation of the parton $i$ appear only though the two functions 
$\gamma_K^{(i)}$ and $\delta_{{\cal J}_i}$. Moreover, the non-Abelian exponentiation theorem \cite{Gatheral:1983cz} implies that the colour structure of these functions is `maximally non-Abelian'. Up to three loops, this implies, in particular (see e.g. Refs.~\cite{Korchemsky:1987wg,Korchemsky:1988si}) that the cusp anomalous dimension depends on the representation only through an overall multiplicative factor, the total colour charge, given by the quadratic Casimir $C_i$ in the representation of parton $i$,
\beq
\label{quad_Casimir}
C_{i} \, {\bf \mathrm{I}} =\sum_a {\rm T}_{i}^{(a)} \, 
{\rm T}_{i}^{(a)} \, ,
\eeq
where ${\bf \mathrm{I}}$ is the unit matrix and ${\bf \rm T}_{i}^{(a)}$ is
a generator in the corresponding representation\footnote{${\rm T}^{(a)}$ 
should be interpreted as follows: for a final--state quark or an initial--state
antiquark: $t^a_{\alpha \beta}$; for a final--state antiquark or an initial--state
quark: $- t^a_{\beta \alpha}$; for a gluon: ${\rm i} \, f_{cab}$. For
${\rm SU}(N_c)$ the index $a$ runs from 1 to $N_c^2 - 1$,  and specifically
$C_F = T_R  (N_c^2 - 1)/N_c$ for quarks and $C_A = N_c$ for gluons. In our 
normalization $T_R = 1/2$.}.
Casimir scaling, namely the universality of $\gamma_K^{(i)} \left( \as \right) / C_{i}$ between quarks and gluons, has been explicitly verified in recent years by three-loop calculation of the QCD splitting functions in Ref.~\cite{Moch:2004pa}. Starting at four loops, however, the colour structure in the exponent may not be expressible in terms of quadratic Casimirs. 
In general, higher Casimir contributions do appear in QCD calculations at this order, for example in the QCD beta function~\cite{vanRitbergen:1997va,Czakon:2004bu}, where one finds colour-singlet contributions constructed of traces of products of four generators. 
The potential appearance of such higher Casimir terms in the exponent, despite the non-Abelian exponentiation theorem \cite{Gatheral:1983cz}, was first observed in Ref.~\cite{Frenkel:1984pz}, where it was proposed to describe the colour structure of the exponent by `colour connected webs', giving a more precise meaning to the notion `maximally non-Abelian'. Recently it has been argued~\cite{Alday:2007mf}, based on different theoretical considerations\footnote{We thank Juan Maldacena for pointing this out to us.}, that such terms may indeed appear in the cusp anomalous dimension starting at four loops.  

Let us therefore write in full generality,
\beq
\label{gamma_K}
\gamma_K^{(i)} \left( \as \right) \equiv C_{i} \, 
\, \widehat{\gamma}_K \left( \as \right) + \widetilde{\gamma}_K^{(i)} \, ,
\eeq
where $C_i$ is given by (\ref{quad_Casimir}), $\widehat{\gamma}_K \left( \as \right) =  2 \, \as/\pi + \ldots$, and $\widetilde{\gamma}_K^{(i)} = {\cal O} (\alpha_s^4)$. 
Note that $\widehat{\gamma}_K \left( \as \right)$ is a universal function of 
the coupling, strictly independent of the representation of the parton~$i$. This
function is known~\cite{Moch:2004pa} up to three loops in QCD. 
In contrast, the residual term $\widetilde{\gamma}_K^{(i)}$ represents (yet unknown)
potential contributions which violate Casimir scaling; it depends
on the representation in a more complicated way, for example through terms that involve irreducible combinations of four colour generators. 
The particular way in which the cusp anomalous dimension depends on the representation will not be important for most of what follows, but it will be used in \secn{ansa} for constructing 
an explicit expression for the soft anomalous dimension.

In a similar way one expects that $\delta_{{\cal J}_i}$ of \eq{G_calJ} would be proportional to the quadratic Casimir at least up to three loops, so we write
\beq
\label{delta_calJ_1loop}
\delta_{\cal J}^{(i)} \left( \as \right)  \equiv C_{i} \, 
\, \widehat{\delta}_{\cal J} \left( \as \right) + \widetilde{\delta}_{\cal 
J}^{(i)} \, ,
\eeq
where $\widehat{\delta}_{\cal J} \left( \as \right) = \as/\pi + \ldots$, 
$\widetilde{\delta}_{\cal J}^{(i)} = {\cal O}(\alpha_s^4)$, and
$C_{i}$ is the Casimir operator defined in \eq{quad_Casimir}. 
The one-loop result quoted here can be deduced from the calculation 
in Ref.~\cite{Dixon:2008gr}.

\subsection{Factorization constraints for soft anomalous dimension matrices}
\label{const}

Having established the simple result in \eq{fincalJ} for the eikonal jet, where 
the kinematic dependence is determined by the cusp anomalous 
dimension, one may wonder if the same is true for the soft function. In other 
words, one may ask whether the full dependence of $\Gamma^{{\cal S}}_{I J}
\left( \beta_i \cdot \beta_j, \as(\mu^2), \e \right)$ in \eq{renS} on 
$\beta_i\cdot\beta_j$ is associated with the cusps. 

One may try to apply the rescaling argument, arguing that if not for the
cusp singularities $\Gamma^{{\cal S}}_{I J}$ should have been invariant with respect to independent rescalings of each~$\beta_i$. One realises however that  
if the number of hard external lines is $n \geq 4$, it is possible to construct
homogeneous conformal cross--ratios such as 
\beq
\label{conformal_ratio_beta}
\rho_{i j k l} \equiv
 \frac{(\beta_i \cdot \beta_j) (\beta_k 
\cdot \beta_l)}{(\beta_i \cdot \beta_k) (\beta_j \cdot \beta_l)}
\eeq
which are inherently invariant with respect to rescalings of each individual 
velocity, thus evading this argument. Kinematic dependence does not necessarily 
lead to violation of the rescaling--invariance property, and therefore might not 
be associated with the cusp singularities.  It is important to note, though, that 
for $n = 2,3$ there are no such ratios and the argument does hold.
One should expect, therefore, that at least for $n = 2,3$ the full kinematic
dependence should be controlled by $\gamma_K$ to all orders. We shall see 
below that this is indeed the case.

The observation that allows us to make a step forward is that the soft 
function, for any number of legs, can be indirectly constrained by considering 
the kinematic dependence of the reduced soft function, defined in \eq{reduS}. 
Here the cusp singularity itself cancels out, and yet, as we will see, it leaves its 
trace through the dependence on the kinematics. To proceed, consider the
renormalization group equation for the reduced soft matrix 
$\overline{{\cal S}}$, which reads
\beq
\label{renbarS}
\mu  \frac{d}{d \mu} \overline{{\cal S}}_{I K}^{[f]} 
\left( \rho_{i j}, \as(\mu^2), \e \right) = - \, 
\Gamma^{\overline{{\cal S}}}_{I J} \left( \rho_{i j}, 
\as(\mu^2) \right) \, \overline{{\cal S}}_{J K}^{[f]} 
\left( \rho_{i j}, \as(\mu^2), \e \right) \, ,
\eeq
where $\Gamma^{\overline{{\cal S}}}_{I J}$, in contrast to 
$\Gamma^{{\cal S}}_{I J}$ and $\gamma_{\cal J}$, is free of singularities.
Its invariance  with respect to scaling of each individual velocity $\beta_i$
is manifest in its functional dependence on the velocities only through the 
ratios $\rho_{ij}$, defined in \eq{rhoij}.

Given the definition of the reduced soft matrix in \eq{reduS}, one easily sees 
that the various eikonal anomalous dimensions are related by
\beq
\label{tran}
\Gamma^{\overline{{\cal S}}}_{I J} \left(\rho_{i j}, \as \right)
= \Gamma^{{\cal S}}_{I J} \left( \beta_i \cdot \beta_j, \as, \e 
\right) - \delta_{I J} \sum_{k = 1}^n \gamma_{{\cal J}_{k}} \left(w_k, 
\as, \e \right) \, ,
\eeq
where, as above, $w_k \equiv 2 (\beta_k \cdot n_k)^2/ n_k^2$. In words, 
pole terms must cancel on the right-hand side of \eq{tran}, and the functional
dependence on eikonal vectors must be arranged so as to reconstruct
functions of $\rho_{i j}$ in order to be consistent with the left-hand side. 
Substituting \eq{fingamJ} into \eq{tran} we get an explicit expression for
$\Gamma^{\overline{{\cal S}}}_{I J}$. In terms of the $D$-dimensional 
running coupling, we can write
\begin{align}
\label{tran_explicit}
\begin{split}
\Gamma^{\overline{{\cal S}}}_{I J} \left(\rho_{i j}, \as(\mu^2,\e) \right) \,
& = \, \Gamma^{{\cal S}}_{I J} \left( \beta_i \cdot \beta_j, \as(\mu^2,\e), 
\e \right) - \delta_{I J} \sum_{k = 1}^n
\bigg[-\frac12 \delta_{{\cal J}_{k}}\left(\as(\mu^2,\e)\right)\, \\&
+ \, \frac14 \gamma_K^{(k)} \left(\as(\mu^2,\e) \right) \, \ln (w_i) \, 
+ \, \frac14 \int_0^{\mu^2} \frac{d \xi^2}{\xi^2} \gamma_K^{(k)} \Big(
\alpha_s (\xi^2,\epsilon) \Big)\bigg]\,.
\end{split}
\end{align}
This immediately implies that
\begin{itemize}
\item{} off-diagonal terms in $\Gamma^{{\cal S}}$ must be \emph{finite}, 
and must depend only on conformal cross ratios of the form of 
$\rho_{ijkl}$ in \eq{conformal_ratio_beta}, which indeed can readily be 
turned into ratios of $\rho_{i j}$'s, as defined in \eq{rhoij}, for example
\beq
\label{conformal_ratio_rho}
\rho_{ijkl} \equiv \frac{(\beta_i \cdot \beta_j) (\beta_k 
\cdot \beta_l)}{(\beta_i \cdot \beta_k) (\beta_j \cdot \beta_l)}
=\left(\frac{\rho_{i j} \, \rho_{k l}}{\rho_{i k} \, \rho_{j l}} \right)^{1/2}
\,{\rm e}^{-{\rm i}\pi(\lambda_{ij} + \lambda_{kl} - \lambda_{ik} -
\lambda_{jl})} \, ;
\eeq
\item{} singular terms in $\Gamma^{{\cal S}}$ must be
confined to diagonal matrix elements, and must be determined by the 
cusp anomalous dimension according to
\beq
\Gamma^{{\cal S}}_{I J} \left( \beta_i \cdot \beta_j, \as(\mu^2, \e), \e 
\right) = \delta_{I J} \sum_{k = 1}^n 
\, \frac14 \int_0^{\mu^2} \frac{d \xi^2}{\xi^2} \gamma_K^{(k)} \Big(
\alpha_s (\xi^2,\epsilon) \Big) + {\cal O}(\epsilon^0) \, ;
\label{singdiag}
\eeq
\item{} finite terms in the diagonal matrix elements must include --- in 
addition to terms that depend exclusively on conformal cross-ratios, as in
\eq{conformal_ratio_rho} --- terms with definite kinematic dependence on 
$\beta_i \cdot \beta_j$, which are proportional to $\gamma_K$, so as to 
properly combine with the $\ln(w_i)$ terms in \eq{tran_explicit}.
\end{itemize}

To illustrate how these features arise, in Appendix \ref{qqbar-one-loop-example} 
we perform an explicit one-loop calculation of a $2 \to 2$ quark-antiquark
scattering amplitude. Note in particular that a given diagram violates rescaling
invariance also in off-diagonal terms, but this violation is eliminated upon 
taking the sum of all diagrams, which is where conformal cross ratios
like $\rho_{ijkl}$ are formed. This is a consequence of gauge invariance.   

Returning to the general case, in Section~\ref{ansa} we will give an explicit 
formula that satisfies the requirements outlined above. Our goal here is to first
formulate the requirements in a compact and general way. To this end, let us
consider the derivative of \eq{tran} (or \eq{tran_explicit}) with respect to 
$\ln (w_i)$. Noting that the $w_i$ dependence appears only through the
eikonal jet functions, and using the results of Section~\ref{eikojet}, which imply
\beq
w_i \, \frac{\partial}{\partial w_i} \, \gamma_{{\cal J}_i} \left( w_i, 
\as, \e \right) = \frac{1}{4} \gamma_K^{(i)} \left( \as \right) \, ,
\label{linlog}
\eeq
we obtain a simple result for the $w_i$-dependence of 
$\Gamma^{{\overline{\cal S}}}_{I J}$,
\beq
w_i \, \frac{\partial}{\partial w_i} \, \Gamma^{{\overline{\cal S}}}_{I J} 
\left( \rho_{i j}, \as \right) = - \, \delta_{I J} \, \, w_i \, 
\frac{\partial}{\partial w_i} \, \gamma_{\cal J} \left( w_i, \as, 
\e \right) \,=\, - \,\frac{1}{4} \gamma_K^{(i)} \left( \as \right) 
\delta_{I J} \, .
\label{linloG}
\eeq
This result can be turned into an equation for the dependence of the 
anomalous dimension matrix on its proper arguments, $\rho_{i j}$, just
using the chain rule. Indeed, for any function $F$ depending on $w_i$ 
only through  $\rho_{i j}$, one finds
\beq
\frac{\partial}{\partial \ln w_i} F \left( \rho_{i j} \right) = 
- \sum_{j \neq i} \frac{\partial}{\partial \ln \rho_{i j}} 
F \left(\rho_{i j} \right) \, .
\label{chain}
\eeq
We conclude that
\beqa
\sum_{j \neq i} \frac{\partial}{\partial \ln \rho_{i j}} 
\Gamma^{{\overline{\cal S}}}_{I J} \left( 
\rho_{i j}, \as \right) & = & 0 \, , \qquad  
\forall i \, , \qquad I \neq J \, , \nonumber \\
\sum_{j \neq i} \frac{\partial}{\partial \ln \rho_{i j}} 
\Gamma^{{\overline{\cal S}}}_{I J} \left( 
\rho_{i j}, \as \right) & = & \frac{1}{4} \, \gamma_K^{(i)} 
\left( \as \right) \, , \qquad  
\forall i \, , \qquad I = J \, .
\label{oureq}
\eeqa
As expected, the first equation in~(\ref{oureq}) states that off-diagonal matrix elements of the soft anomalous dimension matrix should be logarithmic functions 
of homogeneous conformal cross--ratios of $\rho_{i j}$'s, such as 
$\rho_{ijkl}$. For diagonal terms, the second equation in~(\ref{oureq}) states 
that inhomogeneous terms are allowed, but they must be proportional to the cusp 
anomalous dimension in the colour representation of parton $i$. We will 
explore the consequences of these constraints in the following sections, 
beginning with the case of two-parton amplitudes.

\section{Two-parton amplitudes}
\label{sec:two_legs}

In this section we consider in some detail the consequences of the new 
constraints in the simplest case of amplitudes with two hard coloured partons. 
We choose to analyse in \secn{conssud} the special case of the spacelike 
Sudakov form factor of a quark, but the results apply, with minor modifications, 
to any amplitude with two hard partons. In \secn{partojet} we consider the 
partonic jet function, which is an important building block in the factorization 
formula, \eq{facamp}, for any amplitude. We use there the results of 
\secn{conssud} to constrain the kinematic dependence of the partonic jet, 
which is significantly more involved than that of the eikonal jet considered 
above.

\subsection{The case of the Sudakov form factor}
\label{conssud}

Let us consider the implications of the factorization constraints derived 
above on the simplest fixed-angle scattering amplitude, the Sudakov form 
factor. We will see that the constraints of~\eq{oureq} lead to a refinement 
of the results of Ref.~\cite{Dixon:2008gr}, since the kinematic dependence 
of the Sudakov soft function can be explicitly determined in terms of the 
cusp anomalous dimension.

As for any amplitude, our starting point is the factorization formula 
of \eq{facamp}, which here takes the form

\beqa
\label{fafofa}
\Gamma \left( \frac{Q^2}{\mu^2}, \alpha_s(\mu^2), \epsilon \right) & = &
H \left( \frac{Q^2}{\mu^2}, \frac{(2p_i \cdot n_i)^2}{n_i^2 \mu^2}, 
\alpha_s (\mu^2) \right) \times {\cal S} \left( \beta_1 \cdot
\beta_2, \alpha_s (\mu^2), \epsilon \right) \nonumber \\
&&  \times \, \, \prod_{i = 1}^2  \frac{{\displaystyle 
J \left( \frac{(2p_i \cdot 
n_i)^2}{n_i^2 \mu^2}, \alpha_s (\mu^2), \epsilon \right)}}{{\displaystyle
{\cal J} 
\left(\frac{2(\beta_i \cdot n_i)^2}{n_i^2}, \alpha_s (\mu^2), \epsilon
\right)}} \, ,
\eeqa
where $Q^2 = (p_1 + p_2)^2 =  2 p_1 \cdot p_2$. For definiteness, we will consider the space-like form factor, $Q^2\,<\,0$.

In the case of the form factor, the soft function ${\cal S}$ is simply the 
eikonal correlator defined by two Wilson lines running along the classical 
light-like parton trajectories, with velocities given by $\beta_1$ and 
$\beta_2$. Thus
\beq
{\cal S} \left( \beta_1 \cdot \beta_2, \alpha_s (\mu^2), \epsilon \right) =
\left<0| \Phi_{\beta_2} (\infty, 0) \Phi_{\beta_1} (0, -\infty) |0\right> \, .
\label{sudasoft}
\eeq
To determine the kinematic dependence of ${\cal S}$, we consider the 
reduced soft function ${\overline{\cal S}}$, which is given by
\beq
{\overline{\cal S}} \left(\rho_{12}, \alpha_s (\mu^2), \epsilon \right) =
\frac{ {\cal S} \left( \beta_1 \cdot \beta_2, \alpha_s (\mu^2), \epsilon 
\right)}{\Pi_{i = 1}^{2} \, {\cal J} \left( w_i, \alpha_s (\mu^2), 
\epsilon \right)} \, ,
\label{Sbardef}
\eeq
where, as before, $w_i \equiv 2 (\beta_i \cdot n_i)^2 /{n_i^2}$ and
$\rho_{12} =(\beta_1 \cdot \beta_2)^2/(w_1 w_2)$; the latter is specific 
to the spacelike momentum configuration where $\lambda_{12}=0$ so the 
phase in \eq{rhoij} is absent. 

The reduced soft function ${\overline{\cal S}}$ obeys the renormalization 
group equation
\beq
\frac{d \ln {\overline{\cal S}} \left(\rho_{12}, \alpha_s(\mu^2), 
\epsilon \right)}{d \ln \mu} 
= - \, \gamma_{{\overline{\cal S}}} \Big(\rho_{12}, 
\alpha_s( \mu^2) \Big) \, ,
\label{Sbareq}
\eeq
which leads to exponentiation. Since ${\overline{\cal S}}$ is a pure counterterm,
one simply gets 
\beq
{\overline{\cal S}} \left(\rho_{12}, \alpha_s(\mu^2), \epsilon \right) =
\exp \left\{ - \frac12  \int_0^{\mu^2}  \frac{d \lambda^2}{\lambda^2}
\, \gamma_{{\overline{\cal S}}} \Big(\rho_{12}, \alpha_s(\lambda^2, 
\epsilon) \Big) \right\} \, ,
\label{Sbar}
\eeq
in analogy with~\eq{solrencalJ} for the eikonal jet.

Factorization now requires the anomalous dimension 
$\gamma_{{\overline{\cal S}}}$ to be a linear function of $\ln \rho_{12}$. 
Indeed, \eq{oureq} in this case reads
\beq
\frac{\partial \gamma_{{\overline{\cal S}}} \left(\rho_{12}, \alpha_s \right)}{\partial \ln \rho_{12}} = \frac14 \gamma_K (\alpha_s) \, ,
\label{sudaoureq}
\eeq
which integrates to
\beq
\gamma_{{\overline{\cal S}}} \left(\rho_{12}, \alpha_s \right) =
\frac14 \gamma_K (\alpha_s ) \, \ln \rho_{12} +
\delta_{{\overline{\cal S}}} \left( \alpha_s \right) \, ,
\label{gamsbarres}
\eeq
where $\delta_{{\overline{\cal S}}} (\alpha_s)$ is introduced as a 
constant of integration, and does not depend on $\rho_{12}$. As expected, 
the dependence of $\gamma_{{\overline{\cal S}}}$ on the kinematic variable 
$\rho_{12}$ is very simple, and is fully determined by the cusp 
anomalous dimension.

Note that, similarly to what was done for the jet function in 
\eq{delta_calJ_1loop}, we may extract from the anomalous 
dimension $\delta_{{\overline{\cal S}}}$ a factor of the Casimir 
operator of the relevant representation, defining
\beq
\label{delta_bar_calS_1loop}
\delta_{{\overline{\cal S}}}^{(i)} \left( \as \right) \equiv C_{i} \, 
\, \widehat{\delta}_{\cal S} \left( \as \right) + 
\widetilde{\delta}_{{\overline{\cal S}}}^{(i)} \, ,
\eeq
where as usual $\widehat{\delta}_{{\overline{\cal S}}} \left( \as \right) = 
\as/\pi + \ldots$, $\widetilde{\delta}_{{\overline{\cal S}}}^{(i)} = 
{\cal O}(\alpha_s^4)$, and $\widehat{\delta}_{{\overline{\cal S}}}$ is a 
universal function of the coupling, independent of the colour representation. 
The one-loop result quoted here will be determined in \eq{olodelta} below.

Using \eq{gamsbarres} and \eq{Sbar}, we can now write down an explicit 
expression for ${\overline{\cal S}}$, where the kinematic dependence is 
completely solved. We find
\beq
{\overline{\cal S}} \left(\rho_{12}, \alpha_s(\mu^2), \epsilon \right) =
\exp \left\{ - \frac{\ln \left( \rho_{12} \right)}{8} \, \int_0^{\mu^2} \frac{d \lambda^2}{\lambda^2} \, \gamma_K \Big( \alpha_s (\lambda^2,\epsilon) 
\Big) \, - \frac12 \int_0^{\mu^2} \frac{d \lambda^2}{\lambda^2}
\delta_{{\overline{\cal S}}} \Big( \alpha_s (\lambda^2, \epsilon) \Big)
\right\} \, .
\label{Sbarfin}
\eeq
Finally, using the definition of ${\overline{\cal S}}$ in \eq{Sbardef} we
obtain an explicit result for the original soft function ${\cal S}$,
\begin{align}
\label{simsoft}
\begin{split}
& {\cal S} \left( \beta_1 \cdot \beta_2, \alpha_s (\mu^2), \epsilon \right)
=  {\overline{\cal S}} \left(\rho_{12}, \alpha_s (\mu^2), \epsilon \right) 
\times{\Pi_{i = 1}^{2} \, {\cal J} \left( w_i, \alpha_s (\mu^2), 
\epsilon \right)} \\ 
& =  \exp \left\{
\frac12 \int_{0}^{\mu^2} \frac{d \xi^2}{\xi^2} \left[
\delta_{\cal S} \Big( \alpha_s (\xi^2, \epsilon ) \Big) 
- \frac12 \gamma_K \Big(\alpha_s (\xi^2, \epsilon) \Big) \,
\ln \left(\frac{- \beta_1 \cdot \beta_2 \mu^2}{\xi^2} \right) \right]
\right\} \, ,
\end{split}
\end{align}
where we defined
\beq
\delta_{\cal S} \left( \alpha_s \right) =
\delta_{\cal J} \left( \alpha_s \right)
- \delta_{{\overline{\cal S}}} \left( \alpha_s \right)\,, 
\label{delta_calS_calJ_relation}
\eeq
combining the two constants of integration introduced in Eqs. (\ref{G_calJ}) 
and (\ref{gamsbarres}). \eq{simsoft} is intuitively appealing: the spacelike or
timelike nature of the eikonal form factor is associated with the explicit 
logarithm multiplying the cusp anomalous dimension, just as is the case for 
the full form factor.

We conclude this section by briefly comparing our results with those reported 
in Ref.~\cite{Dixon:2008gr}. In order to do so, consider the anomalous 
dimension of the soft function ${\cal S}$, defined by
\beq
\label{renseik}
\frac{d \ln {\cal S} \left( \beta_1 \cdot \beta_2, \alpha_s, 
\epsilon \right)}{d \ln \mu}  = 
- \, \gamma_{\cal S} \left(\beta_1 \cdot \beta_2, 
\as, \e \right) 
\eeq
Using \eq{simsoft}, we obtain an explicit expression for $\gamma_{\cal S}$. 
At the scale $\mu^2$, it reads
\beqa
\label{gamma_S_Sud_final}
\gamma_{\cal S} \left(\beta_1 \cdot \beta_2, 
\as(\mu^2,\e), \e \right) &=&- \, \delta_{\cal S} \Big(\alpha_s(\mu^2,\e) 
\Big) + \frac12 \gamma_K \Big( \alpha_s (\mu^2,\e) \Big) \,
\ln \left( - \beta_1 \cdot \beta_2 \right)
\,\nonumber \\
&& + \, \frac12 \int_0^{\mu^2} \frac{d \xi^2}{\xi^2}
\gamma_K \left(\alpha_s (\xi^2, \epsilon) \right) \, ,
\eeqa
which is analogous to the result for the anomalous dimension of the eikonal 
jet in \eq{fingamJ}.  One may of course verify the consistency of the various renormalization group equations corresponding to \eq{Sbardef}, observing that
\beq
\gamma_{{\overline{\cal S}}} \left(\rho_{12}, \alpha_s \right) =
\gamma_{\cal S} \left(\beta_1 \cdot \beta_2, \as, \e \right)
-\gamma_{{\cal J}_{1}} \left( w_1, \as, \e \right)
-\gamma_{{\cal J}_{2}} \left( w_2, \as, \e \right)\,,
\label{checkcons}
\eeq
where logarithms of different arguments on the right-hand side nicely combine 
to form a logarithm of the scale--invariant ratio $\rho_{12}$, as expected.

We can compare our final result for $\gamma_{\cal S}$ in 
\eq{gamma_S_Sud_final} to the result of Ref.~\cite{Dixon:2008gr}, 
where the same anomalous dimension is written as
\beq
\label{DMS_gamma_calS_Sud}
\gamma_{\cal S} \left(\beta_1 \cdot \beta_2, 
\as(\mu^2,\e), \e \right) 
\\ =  - \, G_{\rm eik} \left( \beta_1 \cdot \beta_2, 
\alpha_s (\mu^2,\e) \right) 
\,+\, \frac12 \int_0^{\mu^2} \frac{d \xi^2}{\xi^2}
\gamma_K \left(\alpha_s (\xi^2, \epsilon) \right) \, ,
\eeq
defining the eikonal function $G_{\rm eik}$. This allows us to relate
$G_{\rm eik}$ to $\delta_{\cal S}$, while solving the dependence
of $G_{\rm eik}$ on the kinematical vectors $\beta_i$. We find
\beq
G_{\rm eik} \left( \beta_1 \cdot \beta_2, \alpha_s \right) =
- \frac12 \, \gamma_K ( \alpha_s ) \,
\ln \left( - \beta_1 \cdot \beta_2 \right)
+ \delta_{\cal S} (\alpha_s ) \, .
\label{Geikform}
\eeq
Finally, let us collect the one-loop expressions for the different eikonal functions
discussed here\footnote{We choose to work with the quark form factor; in case 
of the gluon one the overall colour factor $C_F$ should simply be replaced by 
$C_A$.}, showing that the kinematic dependence found in explicit calculations is 
consistent with the general statements we have made. The soft function 
${\cal S}$ was computed at one loop in~\cite{Botts:1989kf}, with the result
\beq
{\cal S} \left( \beta_1 \cdot \beta_2, \as, \epsilon \right)
= 1 - \frac{\alpha_s}{4 \pi} C_F \left[ \frac{2}{\epsilon^2} - \frac{2}{\epsilon}
\ln(- \beta_1 \cdot \beta_2) \right] \, +\,{\cal O}(\as^2) \, .
\label{olosudasoft}
\eeq
The eikonal jet at this order can be computed by combining \eq{fincalJ}
with \eq{delta_calJ_1loop}, obtaining
\beq
  {\cal J}_i \left(w_i, \as,\e \right) = 1 - \, \frac{\alpha_s}{4 \pi} \, C_F 
  \left[ \frac{1}{\e^2} + \frac{1}{\e} \Big( 1 - \ln \left( w_i \right) \Big)
  \right]\, +\,{\cal O}(\as^2) \, .
\label{olosudaejet}
\eeq
According to \eq{Sbardef}, the reduced soft function is then given by
\beq
{\overline{\cal S}}\left(\rho_{12}, \as, \epsilon \right) = 1 +
\frac{\as}{2 \pi} \, C_F \, \frac{1}{\e} \left[ 1 + \frac{1}{2} 
\ln (\rho_{12}) \right] \,  +\,{\cal O}(\as^2) \, ,
\label{olosudaredusoft}
\eeq
which is indeed a function of $\rho_{12}$, consistent with the general
expression in \eq{Sbarfin}. Taking the logarithmic derivative of 
\eq{olosudaredusoft} one computes the anomalous dimension
\beq
\gamma_{{\overline{\cal S}}}(\rho_{12},\as ) =
\frac{\alpha_s}{\pi} \, C_F \, \left(1 + \frac12 \ln(\rho_{12}) 
\right) \, +\,{\cal O}(\as^2) \, .
\label{gammaSbar}
\eeq
The $\rho_{12}$ dependence of $\gamma_{{\overline{\cal S}}}$, as expected, 
is given by \eq{gamsbarres}, when the one-loop result for 
$\gamma_K^{(1)} = 2 \, C_F \, \as/\pi$ is used. The leading-order term of 
the function $\delta_{{\overline{\cal S}}}$ is then given by
\beq
\delta_{{\overline{\cal S}}} (\as) \, = \, 
 C_F\frac{\alpha_s}{\pi} \, +\,{\cal O}(\as^2) \, ,
\label{olodelta}
\eeq
while, using \eq{delta_calS_calJ_relation} and \eq{delta_calJ_1loop}, we 
verify that $\delta_{\cal S} = {\cal O}(\as^2)$, in agreement with the results 
of Ref.~\cite{Dixon:2008gr}.

To summarize, we have shown that the kinematic dependence of all purely 
eikonal functions entering the form factor can be reconstructed using
constraints arising from factorization, together with general properties of pure counterterms in dimensional regularization. The result is that the entire kinematic dependence of these functions is proportional to the cusp anomalous dimension. This confirms our initial expectation, based on the fact that violation of rescaling invariance in such functions can only be introduced by the cusp anomaly. 
In \secn{ansa} we will return to the generic case of multi-leg fixed-angle 
scattering amplitudes, and work out the consequences of our constraints. 
Before that, however, let us briefly contrast our findings concerning eikonal
functions with the case of partonic amplitudes. Again, we choose the simplest example, that of the partonic jet. 

\subsection{Constraining the partonic jet function}
\label{partojet}

We have seen that the kinematic dependence of eikonal functions is remarkably
simple, to all-order in perturbation theory. The singularity structure of a partonic
amplitude resembles that of the corresponding eikonal function. Nevertheless,
because partonic amplitudes do depend on dimensionful kinematic variables and
have infrared singularities as well as distinct ultraviolet renormalization properties,
they do not admit similarly simple all-order expressions.  

An exception is the Sudakov form factor $\Gamma(Q^2,\epsilon)$ itself, which
does not get renormalized ($d \Gamma(Q^2,\epsilon)/\ln \mu=0$), so that its
entire perturbative expansion in dimensional regularization is determined by its
infrared singularities~\cite{Magnea:1990zb,Dixon:2008gr}, according to
\beq
\label{Sud_FF_G_plus_K}
\Gamma(Q^2,\e)=\exp\left\{\frac12 \int_{0}^{-Q^2}\frac{d\xi^2}{\xi^2}
\left[G\left(-1,\as(\xi^2,\e),\e\right) -\frac12 
\gamma_K\left(\as(\xi^2,\e)\right)
\ln \left(\frac{-Q^2}{\xi^2}\right)\right]\right\}\,.
\eeq
Indeed, in this case the eikonal function in \eq{simsoft} is very similar to the full
form factor, and apart from the different scales, the only qualitative difference 
is the appearance of non-negative powers of $\epsilon$ in the function 
$G\left(-1,\as,\e\right)$.

In a general scattering amplitude, which does get renormalized, the dependence 
on the renormalization point $\mu^2$ and on the kinematic variables is of course
distinct, and neither of them is associated exclusively with the cusp anomalous 
dimension. The simplest example is that of the partonic jet, which we now consider.

Using the factorization formula for the Sudakov form factor, \eq{fafofa}, 
together with Eqs. (\ref{Sud_FF_G_plus_K}), (\ref{fincalJ}) and (\ref{simsoft}), 
one can directly determine the product
\beqa
H \left( \frac{Q^2}{\mu^2}, \frac{(2 p_i \cdot n_i)^2}{n_i^2 \mu^2}, 
\alpha_s \right) \prod_{i = 1}^2 J \left( \frac{(2 p_i \cdot n_i)^2}{n_i^2 
\mu^2}, \alpha_s, \e \right) \equiv \, \prod_{i = 1}^2 \widetilde{J} 
\left( \frac{(2 p_i \cdot n_i)^2}{n_i^2 \mu^2}, \alpha_s, \e\right)\,,
\eeqa
where we defined 
\begin{align}
\label{tilde_J}
\begin{split}
&\widetilde{J}\left( \frac{(2 p_i \cdot n_i)^2}{n_i^2 \mu^2}, \alpha_s, 
\e \right) = \\
& \hspace*{30pt} \exp \bigg\{
\frac14 \int_{\mu^2}^{- Q^2}\frac{d \lambda^2}{\lambda^2} \left[
- \frac12 \gamma_K \left( \alpha_s (\lambda^2, \e) \right) \,
\ln \left( \frac{- Q^2}{\lambda^2} \right)
\, + \, G (- 1, \alpha_s (\lambda^2, \e), \e)
\right] \\
&\hspace*{45pt} + \frac14 \int_0^{\mu^2} \frac{d\lambda^2}{\lambda^2}
\left[- \frac12 \gamma_K \left( \alpha_s (\lambda^2, \e) \right)
\ln \left(\frac{(2 p_i \cdot n_i)^2}{n_i^2 \lambda^2} \right)
\, + \, \delta_J ( \alpha_s (\lambda^2, \epsilon), \epsilon) \right]
\bigg\} \, ,
\end{split}
\end{align}
with
\begin{align}
\label{delta_J_def}
\delta_J(\alpha_s, \epsilon) \equiv
\delta_{{\overline{\cal S}}} (\alpha_s) + G(- 1, \alpha_s, \epsilon) \, 
\end{align}
where $\delta_{{\overline{\cal S}}}$ is the function introduced in 
\eq{gamsbarres}. Note that $\delta_J(\alpha_s,\epsilon)$, in contrast 
to $\delta_{\cal J}(\alpha_s)$ in \eq{G_calJ}, depends on $\e$ explicitly, 
and it does have non-negative powers of $\epsilon$ coming from 
$G(-1,\alpha_s ,\epsilon)$.

We now observe that in \eq{tilde_J} all infrared singularities emerge from the 
$\lambda^2 \to 0$ limit of the integral over the running coupling in the last line.
These singularities must all reside in the partonic jet function $J$, while the renormalized hard coefficient function $H$ is finite. This observation, together 
with the renormalization group equation for $J$, \eq{renJ}, and the fact that 
$J$ does not depend explicitly on $Q^2$, implies that $J$ takes the form
\beqa
\label{final_J}
J \left(\frac{p_n^2}{\mu^2}, \alpha_s(\mu^2), \epsilon \right) 
& = & \, \exp \Bigg\{ h_J \left( \alpha_s (p_n^2) \right)
+ \frac12 \int_{\mu^2}^{p_n^2} \frac{d\lambda^2}{\lambda^2} 
\gamma_J \left( \alpha_s(\lambda^2) \right) \\
& &\hspace{10pt} + \int_0^{p_n^2} \frac{d \lambda^2}{\lambda^2} 
\left[- \frac18 \gamma_K \left( \alpha_s( \lambda^2, \epsilon) \right)
\ln \left( \frac{p_n^2}{\lambda^2} \right) + \frac14 \delta_J \left(
\alpha_s(\lambda^2, \epsilon), \epsilon \right) \right] \Bigg\} \, ,
\nonumber
\eeqa
where we defined $p_n^2\equiv (2 p\cdot n)^2/n^2$ for brevity. All the 
functions appearing here are finite as $\epsilon \to 0$, so also here singularities 
are generated only through the integration over the running coupling in the last 
line. The function $\delta_J ( \alpha_s (\lambda^2), \epsilon )$  depends on 
$\epsilon$ explicitly, while $\gamma_K$, $\gamma_J$ and $h_J$ do not. 
As far as $h_J$ is concerned, this last statement is not obvious a priori, and 
it will be proven to all orders below.

The structure of the result in \eq{final_J} for the partonic jet is intuitively 
clear: infrared and collinear singularities are generated by the integration over 
the scale of the running coupling in the second line. These terms in the 
exponent are similar to the expression found for the eikonal jet, \eq{fincalJ}, 
and indeed the double poles of the two expressions match, as they must. 
Single poles, on the other hand, are different, on account of hard collinear 
radiation, which is not correctly approximated by ${\cal J}$. Further dependence 
on the hard scale $p_n^2$ arises in the first line of \eq{final_J}, due to the 
non-trivial ultraviolet behaviour of $J$, which is dictated by $\gamma_J$. It 
is natural in this case to start the renormalization group evolution at $\mu^2 
= p_n^2$, where the function $h_J$ gives the initial condition.

Once again, we may compare our all-order expression with one-loop results.
One may start with the well-known result~\cite{Magnea:1990zb} for the function
$G$,
\beq
G \left( \frac{Q^2}{\mu^2}, \as, \e \right)
= \frac{\alpha_s}{\pi} \, C_F \,\left[
\left( \frac{\mu^2}{- Q^2}\right)^{\e} \left(\frac{1}{\e} +
\frac32 - {\e} \left(\frac{\pi^2}{12} - 4 \right) + 
{\cal O}(\e^2)\right) - \frac{1}{\e} \right]
+ {\cal O}(\as^2) \, .
\label{oldG}
\eeq
Using \eq{delta_J_def} and \eq{gamsbarres} one then finds
\beq
\label{delta_J_1loop}
\delta_J \left( \alpha_s, \epsilon \right) =
\frac{\alpha_s}{\pi} \, C_F \,\left(\frac52 \,-\,{\e}\left(\frac{\pi^2}{12}-4\right)\,+\,{\cal O}(\e^2)\right) 
\,+\,{\cal O}(\as^2) \, .
\eeq
The renormalization of the jet, on the other hand, yields
\beq
\label{gamma_J_1loop}
\gamma_J (\as) = - \frac34 \, C_F \, \frac{\alpha_s}{\pi} \,+\,
{\cal O}(\as^2) \, .
\eeq
Inserting the one-loop results of Eqs. (\ref{gamma_K}), (\ref{delta_J_1loop}) 
and (\ref{gamma_J_1loop}) into our general expression, \eq{final_J}, and
expanding to ${\cal O}(\alpha_s)$, we obtain finally
\begin{align}
\label{final_J_one_loop}
\begin{split}
J \left( \frac{p_n^2}{\mu^2}, \alpha_s(\mu^2), \epsilon \right)
& = 1 + h_J^{(1)} (\alpha_s) +  \frac{\alpha_s}{4\pi} \, C_F \,
\bigg\{ - \frac{1}{\e^2} - \frac{1}{\e} \left[ \frac{5}{2} - \ln 
\left( \frac{p_n^2}{\mu^2} \right) \right] \\
& \hspace*{-40pt} - \frac12 \ln^2 \left( \frac{p_n^2}{\mu^2} \right) +
\ln \left( \frac{p_n^2}{\mu^2} \right) + 4 - \frac{\pi^2}{12}
+ {\cal O}(\e) \bigg\} \, + \, {\cal O}(\as^2) \, .
\end{split}
\end{align}
Comparing this result to the explicit one-loop calculation, reported in Section 3 of Ref.~\cite{Dixon:2008gr}, we find full consistency and determine the finite coefficient $h_J$,
\beq
h_J (\alpha_s) = - \frac{\alpha_s}{\pi} \, C_F \, \left( \frac32 +
\frac{\pi^2}{12} \right) \, + \, {\cal O}(\as^2) \, .
\label{hJ1}
\eeq
We conclude this section by comparing our parametrization of the partonic 
jet, \eq{final_J}, with the results of Ref.~\cite{Dixon:2008gr}. This will then 
be used to relate the function $h_J (\alpha_s)$, which sets the normalization 
of the jet function to the hard function appearing in the factorized the Sudakov
form factor.  We begin by considering the differential equation that controls 
the dependence of $J$ on $p_n^2$. This is, once again, a `$K + G$' equation, 
and can be written as~\cite{Dixon:2008gr}
\beq
\label{calG_def}
\frac{\partial \ln J \left(p_n^2/\mu^2, \alpha_s (\mu^2), \epsilon 
\right)}{\partial \ln p_n^2} = \frac12 \, {\cal G} \left( \frac{p_n^2}{\mu^2}, 
\alpha_s(\mu^2), \e \right) - \frac18 \int_{0}^{\mu^2} 
\frac{d \lambda^2}{\lambda^2} \gamma_K
\left( \alpha_s( \lambda^2, \e) \right) \, ,
\eeq
where the cusp--related singularity was explicitly extracted. The renormalization
group equation for $J$, \eq{renJ}, implies that
\beq
\frac{d }{d \ln \mu^2}
\frac{ \partial \ln J \left( p_n^2/\mu^2, \alpha_s, \epsilon \right)}{\partial 
\ln p_n^2} = \frac{\partial}{\partial \ln p_n^2}
\frac{d \ln J \left( p_n^2/\mu^2, \alpha_s, \epsilon \right)}{d \ln 
\mu^2} = 0 \, .
\eeq
Applying this to \eq{calG_def} in turn implies that the scale dependence of 
${\cal G}$ is controlled by the cusp anomalous dimension, according to
\beq
 \frac{d}{d\ln \mu^2} \, {\cal G} \left( \frac{p_n^2}{\mu^2}, 
 \alpha_s (\mu^2), \e \right) = \frac14 \, \gamma_K \left( \as \right) \, ,
 \label{evcalG}
\eeq
or, upon integration,
\beq
{\cal G} \left( \frac{p_n^2}{\mu^2}, \alpha_s(\mu^2), \e \right) = 
{\cal G} \left(1, \alpha_s (p_n^2), \e \right) - \frac14 
\int_{\mu^2}^{p_n^2} \frac{d \lambda^2}{\lambda^2} \gamma_K
\left( \alpha_s( \lambda^2, \e ) \right) \, .
\label{solevcalG}
\eeq

Now, using our general expression for the partonic jet in \eq{final_J}, we 
can readily interpret the function $\cal G$ in (\ref{calG_def}) in terms of 
the anomalous dimensions $\gamma_J$ and $\delta_J$, and
of the finite function $h_J$. We find
\beq
\label{cal_G_result}
{\cal G} \left( 1, \as, \e \right) = \beta (\as, \e) \, 
\frac{d}{d \alpha_s}  h_J ( \alpha_s ) + \gamma_{J}( \alpha_s ) +
\frac12 \delta_J( \alpha_s, \e ) \, ,
\eeq
where the first term is proportional to the $\beta$ function, $\beta(\as, \e)
= {d \alpha_s(\mu^2, \e)}/{d\ln \mu}$, so that it starts contributing at 
${\cal O}(\alpha_s^2)$. 

Using the definition of $\delta_J(\alpha_s, \epsilon)$, given in \eq{delta_J_def},
and setting\footnote{Note that in setting $(2p\cdot n)^2/n^2=\mu^2$ we assume that the auxiliary vector $n$ is timelike, $n^2>0$. A spacelike $n$ is also possible of course, and this choice would be reflected in replacing ${\cal G}(1,\as,\e)$ below by ${\cal G}(-1,\as,\e)$.} $\mu^2 = p_n^2$, we may now obtain
an expression for the function $G$ appearing in the Sudakov form factor, 
\eq{Sud_FF_G_plus_K}, which we can compare to the results of 
Ref.~\cite{Dixon:2008gr}. We find
\beq
\label{G_ours}
G \left(- 1, \as, \epsilon \right) = - 2 \, \beta(\as, \e)
\frac{d}{d\alpha_s} \, h_J(\alpha_s) - \delta_{{\overline{\cal S}}}(\as)
- 2 \, \gamma_{J} (\as) + 2 \, {\cal G}(1,\as,\e) \, .
\eeq
\eq{G_ours} can be compared with the result of Section 4 in 
Ref.~\cite{Dixon:2008gr}, which reads 
\beq
\label{G_DMS}
G( - 1, \as, \epsilon ) = \beta (\alpha_s, \e) \, 
\frac{\partial}{\partial \as} \, \ln H(- 1, w_i, \as) 
- \gamma_{{\overline{\cal S}}} (\rho_{12}, \as) - 2 \gamma_J(\as)
+ \sum_{i = 1}^2 {\cal G}_i (w_i, \as, \e) \, ,
\eeq
where we have chosen a specific common normalization for the velocities, 
$p_i = \beta_i Q_0/\sqrt{2}$, with $Q_0 = \sqrt{- Q^2}$, corresponding 
to $(2 p_i \cdot n_i)^2/n_i^2 = - Q^2 \, w_i$, and $\beta_1 \cdot \beta_2
= -1$, so that $\rho_{12} = 1/(w_1\,w_2)$. Using now our result for
$\gamma_{{\overline{\cal S}}} (\rho_{12}, \as)$, \eq{gamsbarres}, we see 
that \eq{G_DMS} takes the form
\begin{align}
\label{G_DMS_explicit}
\begin{split}
G (- 1, \as, \epsilon) \, = & \, \, \beta( \alpha_s, \e) 
\frac{\partial}{\partial \as} \ln H(- 1, w_i, \as) 
+ \frac{1}{4} \, \gamma_K(\alpha_s) \,
\ln(w_1\, w_2) \\ & - \delta_{{\overline{\cal S}}} (\alpha_s)
- 2 \, \gamma_J (\as) + \sum_{i = 1}^2 {\cal G}_i(w_i, \as, \e) \, .
\end{split}
\end{align}
\eq{G_DMS_explicit} holds for arbitrary $w_1$ and $w_2$, implying that the dependence of the \emph{r.h.s.} on these parameters through the explicit 
$\gamma_K$ term, and through the functions $H$ and ${\cal G}$, 
must cancel out. 

Comparing Eqs.~(\ref{G_DMS_explicit}) and (\ref{G_ours}) we obtain 
\begin{align}
\begin{split}
- 2 \, \beta(\as, \e) \, \frac{d}{d\alpha_s} h_J(\alpha_s) \,
& + 2 \, {\cal G}(1, \as, \e) \, = 
\frac{1}{4} \, \gamma_K(\alpha_s) \, \ln(w_1\,w_2) 
\\ & + \beta(\as, \e) \frac{\partial}{\partial \as} \ln H(- 1, w_i, \as)
+ \sum_{i = 1}^2 {\cal G}_i (w_i, \as, \e) \, .
\end{split}
\end{align}
We already know that the dependence of the \emph{r.h.s.} on the $w_i$ 
cancels out, so we are allowed to use this equality for any $w_i$. Picking 
$w_1 = w_2 = 1$, we obtain
\beq
\label{eta_J_H_relation}
h_J (\alpha_s) = - \frac12 \ln H (- 1, w_i = 1, \as) \, ,
\eeq
which determines the function $h_J$ in terms of the hard function 
$H$ of the Sudakov form factor. \eq{eta_J_H_relation} implies in particular 
that $h_J (\as)$ does not carry any explicit dependence on $\epsilon$, but
depends only on the coupling, as anticipated.

\section{Multi-parton amplitudes: the soft anomalous dimension matrix}
\label{ansa}

Let us now consider the general case of multi-parton amplitudes and examine 
the consequences of our new constraints on the all-order structure of the 
soft anomalous dimension matrix.  We first note that the constraint of 
\eq{oureq}, which can be written as
\beqa
\label{oureq_reformulated}
\sum_{j,\, j \neq i} \frac{\partial}{\partial \ln(\rho_{i j})} \,
\Gamma^{{\overline{\cal S}}}_{MN} \left( 
\rho_{i j}, \as \right) & = & \frac{1}{4} \, \gamma_K^{(i)} 
\left( \as \right) \,\delta_{MN} \, ,\qquad\qquad \forall i\,,
\eeqa
is an equality between colour matrices, valid in any basis. We will then proceed 
to work with the colour generator notation, as in \cite{Catani:1998bh}, without
specifying a basis; consequently, we will drop the explicit matrix indices $M$ 
and $N$ in the following.

Next, we observe that \eq{oureq_reformulated} effectively relates the colour
structure of the soft matrix, which is a priori very complicated, to the much 
simpler colour structure on the right hand side. Clearly the particular way in 
which the cusp anomalous dimension $\gamma_K^{(i)}$  depends on the 
colour representation of parton $i$ becomes important. Following 
\eq{gamma_K}  we can write
\beqa
\label{oureq_reformulated1}
\sum_{j,\, j \neq i} \frac{\partial}{\partial \ln(\rho_{i j})} \,
\Gamma^{{\overline{\cal S}}} \left( 
\rho_{i j}, \as \right) & = & \frac{1}{4} \, \bigg[C_i \,\widehat{\gamma}_K \left( \as \right) \,+\, \widetilde{\gamma}_K^{(i)}\left( \as \right)\bigg] \, ,\qquad\qquad \forall i\,,
\eeqa
where in the first term the dependence on the representation is explicit, while in the second it is implicit. Using the linearity of these equations we can obviously write the general solution as a superposition of two functions
\beq
\Gamma^{{\overline{\cal S}}}=\Gamma^{{\overline{\cal S}}}_{{\text{Q.C.}}}
+\Gamma^{{\overline{\cal S}}}_{{\text{H.C.}}}
\eeq
which are, respectively, solutions of the equations
\begin{align}
\label{oureq_reformulated_QC}
\sum_{j,\,j \neq i} \frac{\partial}{\partial \ln(\rho_{i j})} \,
\Gamma^{{\overline{\cal S}}}_{{\text{Q.C.}}} \left( 
\rho_{i j}, \as \right) & =  \frac{1}{4} \, 
\left(\sum_a {\rm T}_i^{(a)} \,{\rm T}_i^{(a)} \right)\,\widehat{\gamma}_K \left( \as \right) \, ,\qquad\qquad \forall i\,,
\\
\label{oureq_reformulated_HC}
\sum_{j,\, j \neq i} \frac{\partial}{\partial \ln(\rho_{i j})} \,
\Gamma^{{\overline{\cal S}}}_{{\text{H.C.}}} \left( 
\rho_{i j}, \as \right) & =  \frac{1}{4} \,  \widetilde{\gamma}_K^{(i)}\left( \as \right)\,, \qquad\qquad \forall i\,.
\end{align}
Here {\text{Q.C.}} and {\text{H.C.}} stand for Quadratic Casimir and Higher 
(order) Casimir, respectively, reflecting the different group theoretical 
structure of the two contributions . In the first equation we exhibited the dependence on the colour representation of parton $i$, using \eq{quad_Casimir}.
In the following we will focus on determining $\Gamma^{{\overline{\cal 
S}}}_{{\text{Q.C.}}}$ leaving aside $\Gamma^{{\overline{\cal 
S}}}_{{\text{H.C.}}} = {\cal O}(\alpha_s^4)$, which is driven by yet 
unknown corrections to the cusp anomalous dimension, that are not 
proportional to $C_i$. For simplicity of the notation we henceforth drop 
the subscript Q.C. 

A solution for $\Gamma^{{\overline{\cal S}}}$, obeying \eq{oureq_reformulated_QC}, 
is given by
\begin{align}
\begin{split}
\label{ansatz}
\left.\Gamma^{\overline{S}}
\left(\rho_{i j}, \as \right) \right\vert_{\rm ansatz} &= \, - \frac18 \,
\widehat{\gamma}_K\left(\as \right) \sum_{i=1}^n \sum_{j,\, j\neq i} \,
\ln(\rho_{ij}) \, \sum_{a} \mathrm{T}_i^{(a)}  \mathrm{T}_j^{(a)} \, \\
& \hspace*{40pt}+ \, \frac12 \, \widehat{\delta}_{{\overline{\cal S}}} ( \alpha_s )  
\sum_{i=1}^n  \sum_{a} \mathrm{T}_i^{(a)} \mathrm{T}_i^{(a)} \, ,
\end{split}
\end{align} 
where $\widehat{\gamma}_K$ and $\widehat{\delta}_{{\overline{\cal S}}}$ 
were defined in \eq{gamma_K} and in \eq{delta_bar_calS_1loop}, respectively.
Note that in \eq{ansatz} the first term, which couples each pair of partons into a 
colour dipole, carries the entire dependence on kinematics, correlating it with the
colour structure, while the second term is independent of kinematics and is 
proportional to the unit matrix in colour space. Such a term, in a different
factorization scheme, such as the one adopted in Refs.~\cite{Sterman:2002qn,
MertAybat:2006mz}, could be associated with jet functions. In our case, since 
we start with specific operator definitions for the jets, we may find leftover 
colour-diagonal contributions in $\Gamma^{\overline{S}}$. Note also that 
\eq{ansatz}, which is valid for a general $n$-parton amplitude, reduces to 
\eq{gamsbarres} in the $n = 2$ case of the Sudakov form factor. This relation 
was used in determining the second term in \eq{ansatz}, which is obviously 
not constrained by \eq{oureq_reformulated_QC}.

The explicit calculation of Ref.~\cite{MertAybat:2006mz} at two-loops has
established that \eq{ansatz} is the full answer to this order. In particular, the 
soft anomalous dimension does not contain correlations involving generators 
of three different partons, such as $f_{abc}\, \mathrm{T}_i^{(a)} 
\mathrm{T}_j^{(b)} \mathrm{T}_k^{(c)}$ (see \eq{hat_H} below), despite 
the fact that single poles in $\epsilon$ at  two loops are known to contain 
such terms.  Ref.~\cite{MertAybat:2006mz} has demonstrated that these 
terms are generated at two loops only upon expansion of the product of the 
soft and hard functions.

To verify that \eq{ansatz} satisfies \eq{oureq_reformulated_QC}, let us take a
derivative with respect to $\ln (\rho_{ij})$, for specific partons $i$ and $j$, 
with $i\neq j$. We find
\beq
\frac{\partial \Gamma^{\overline{S}} \left( 
\rho_{i j}, \as \right) }{\partial \ln(\rho_{ij})} =
-\frac14 \, \widehat{\gamma}_K \left( \as \right) \,
\sum_a \mathrm{T}_i^{(a)} \mathrm{T}_j^{(a)} \, ,
\eeq 
where we used the fact that $\sum_a \mathrm{T}_j^{(a)} \mathrm{T}_i^{(a)} 
= \sum_a \mathrm{T}_i^{(a)} \mathrm{T}_j^{(a)}$. Next we sum over $j$
for fixed $i$, as in the \emph{l.h.s.} of (\ref{oureq_reformulated_QC}), and we get
\begin{align}
\label{proof}
\begin{split}
\sum_{j,\,j\neq i}\frac{\partial \Gamma^{\overline{S}} \left( 
\rho_{i j}, \as \right)}{\partial \ln(\rho_{ij})} 
& = - \frac14 \widehat{\gamma}_K\left( \as \right) \,\sum_{j,\,j\neq i}
\sum_a \mathrm{T}_i^{(a)} \mathrm{T}_j^{(a)} \\
& = - \frac14 \widehat{\gamma}_K\left( \as  \right) \,
\sum_a \mathrm{T}_i^{(a)}  \left(-\mathrm{T}_i^{(a)}\right) \, ,
\end{split}
\end{align}
which coincides with the \emph{r.h.s.} of \eq{oureq_reformulated_QC}, 
as required. The second line of \eq{proof} follows from colour conservation, 
which is expressed in the colour generator notation simply by 
\begin{align}
\sum_{i = 1}^n \mathrm{T}_i^{(a)} = 0 \, .
\label{T_prop2}
\end{align}
Clearly, our ansatz does not in general provide a unique solution. Indeed, 
\eq{oureq_reformulated_QC} is a set of $n$ linear differential equations in the 
variables $\ln (\rho_{ij})$, while the number of variables is quadratic in $n$, 
so there may be contributions beyond \eq{ansatz}. To clarify the nature of 
possible corrections to our ansatz, let us define
\beq
\label{Delta_def}
\Gamma^{\overline{S}}_{MN} \left( 
\rho_{i j}, \as \right) = \left. \Gamma^{\overline{S}}_{MN} \left( 
\rho_{i j}, \as \right) \right\vert_{\rm ansatz}
+ \Delta^{\overline{S}}_{MN} \left( \rho_{i j}, \as \right)
\eeq
and summarize our knowledge of $\Delta^{\overline{S}}_{MN}$. First, as 
discussed in \secn{const}, and shown explicitly in \secn{conssud} and 
Appendix~\ref{three_partons}, $\Delta^{\overline{S}}_{MN}$ vanishes 
identically for $n = 2, 3$. Thus it may contribute only to $n \geq 4$ parton
amplitudes, starting at three loops. Second, according to \eq{oureq_reformulated_QC}, it should satisfy 
the homogeneous equation:
\beq
\label{Delta_oureq_reformulated}
\sum_{j,\,j \neq i} \frac{\partial}{\partial \ln(\rho_{i j})} 
\Delta^{{\overline{\cal S}}}_{MN} \left( 
\rho_{i j}, \as \right) =  0 \, \qquad \forall i, \, M, \, N\,.
\eeq
\eq{Delta_oureq_reformulated} is solved by any function that depends on 
$\rho_{ij}$ only through conformal cross ratios such as $\rho_{ijkl}$,
defined in \eq{conformal_ratio_beta}\footnote{As explained in \secn{const} 
and illustrated in Appendix \ref{qqbar-one-loop-example}, conformal cross 
ratios appear in $\Gamma^{{\overline{\cal S}}}$ also in (\ref{ansatz}), starting 
at one-loop.}. As an example, for four external partons there are only two
independent conformal cross ratios (note that $1/\rho_{1423} = \rho_{1234}
\, \rho_{1342}$), and  the general solution can be written as
\beq
\Delta^{{\overline{\cal S}}}_{MN} = \Delta^{{\overline{\cal S}}}_{MN}
\left(\rho_{1234}, \rho_{1342}, \as \right) \, .
\eeq
Note that \eq{Delta_oureq_reformulated} does not restrict the functional dependence of $\Delta^{{\overline{\cal S}}}_{MN}$ on conformal cross ratios,
allowing in particular non-linear dependence. Interesting examples, still for four 
partons, are
\beq
\label{hat_H}
\hat{\bf H}^{(2)}_{[\rm f]}= \sum_{j,k,l} \sum_{a,b,c}
{\rm i} \, f_{abc} \,{\rm T}_j^{a} {\rm T}_k^{b} {\rm T}_l^{c} \,
\ln \left(\rho_{ijkl}\right) \, \ln \left(\rho_{iklj}\right) \,
\ln \left(\rho_{iljk}\right) \, ,
\eeq
and
\beq
\widetilde{\bf H}_{[\rm f]}= \sum_{j,k,l} \sum_{a,b,c}
d_{abc} \, {\rm T}_j^{a} {\rm T}_k^{b} {\rm T}_l^{c} \,
\ln^2 \left(\rho_{ijkl}\right) \, \ln^2 \left(\rho_{iklj}\right) \,
\ln^2 \left(\rho_{iljk}\right) \, ,
\eeq
where in both equations the sum over partons is understood to exclude 
identical indices. As already mentioned, $\hat{\bf H}^{(2)}_{[\rm f]}$ is 
known to appear\footnote{In Ref.~\cite{MertAybat:2006mz} 
$\hat{\bf H}^{(2)}_{\rm f}$ is written in terms of the three kinematic 
invariants of a four-leg amplitude, assuming momentum conservation. 
It is straightforward to show that the same expression can be written in 
terms of conformal cross ratios, where momentum conservation is not 
enforced.} as part of the ${\cal O}(1/\epsilon)$ coefficient in two loops 
amplitudes. Nevertheless, as was shown in Ref.~\cite{MertAybat:2006mz}, 
it does not appear in the soft anomalous dimension at this order. 

For multi-leg amplitudes with more than four legs, the space of possible 
solutions to the available constraints increases further, as there is an 
increasing number of conformal cross ratios. It remains for future work 
to decide whether contributions beyond the ansatz of \eq{ansatz} do 
indeed appear.

We conclude by working out the consequences of \eq{ansatz} for the 
anomalous dimension of the original soft matrix ${\cal S}$. Using \eq{tran}, 
we write
\beq
\label{tran_in_reverse}
\left. \Gamma^{{\cal S}}_{I J} \left( \beta_i \cdot \beta_j, \as(\mu^2), 
\e \right)\right\vert_{\rm ansatz} \, = \,
\left. \Gamma^{\overline{{\cal S}}}_{I J} \left(\rho_{i j}, \as(\mu^2) \right)
\right\vert_{\rm ansatz} \, + \,
\delta_{I J} \sum_{i = 1}^n \gamma_{{\cal J}_i} \left(w_i, 
\as(\mu^2), \e \right) \, ,
\eeq
where, as before, $w_i = 2 (\beta_i \cdot n_i)^2/n_i^2$. Substituting here
\eq{ansatz}, together with the expression in \eq{fingamJ} for the eikonal jet
anomalous dimension, we obtain, after some algebra
\begin{align}
\begin{split}
\label{ansatz_calS}
\left.\Gamma^{{\cal S}} \left( 
\beta_i \cdot \beta_j, \as(\mu^2), \e \right) \right\vert_{\rm ansatz} 
& = \, - \frac14 \, \widehat{\gamma}_K \left(\as(\mu^2) \right) 
\sum_{i=1}^{n} \sum_{j,\,j \neq i} \, \ln \left(\beta_i \cdot \beta_j \, {\rm e}^{{\rm i} 
\pi \lambda_{ij}} \right) \, \sum_{a} \mathrm{T}_i^{(a)} 
\mathrm{T}_j^{(a)} \, \\
& \hspace*{-40pt} + \, \left[- \frac12 \, \widehat{\delta}_{\cal S}
\left( \alpha_s(\mu^2) \right) + \frac14 \, \int_0^{\mu^2} 
\frac{d \lambda^2}{\lambda^2} \widehat{\gamma}_K 
\left(\as(\lambda^2, \e) \right) \right] \, \sum_{i=1}^n \sum_{a}
\mathrm{T}_i^{(a)}  \mathrm{T}_i^{(a)} \, ,
\end{split}
\end{align} 
where, as in \eq{delta_calS_calJ_relation},
\beq
\widehat{\delta}_{\cal S}(\alpha_s) =\widehat{\delta}_{\cal J}(\alpha_s)\,-\,\widehat{\delta}_{{\overline{\cal S}}}(\alpha_s)\,.
\eeq
The manipulation we applied in order to write \eq{tran_in_reverse} in its 
final form is similar to the one we used in \eq{proof}, but in the reverse 
order: here one first rewrites the colour--diagonal terms that depend on 
$\ln (w_i)$, which originate in the eikonal jet contributions, in terms of a 
sum over all other partons $j$, using $\mathrm{T}_i^{(a)} = - \sum_{j 
\neq i} \mathrm{T}_j^{(a)}$; then one combines these terms with 
the $\ln (\rho_{ij})$ terms in $\Gamma^{{\overline{\cal S}}}$; finally, using
\eq{rhoij}, one observes that $\Gamma^{{\cal S}}$ depends only on 
$\beta_i\cdot \beta_j$, as it must. It is straightforward to check that 
\eq{ansatz_calS} reduces to \eq{gamma_S_Sud_final} for the Sudakov 
form factor. Once again, the second line in \eq{ansatz_calS} is proportional
to the identity matrix in colour space, and could be associated with jets
in a different factorization scheme. Indeed, the square bracket gives just 
one half of the kinematics-independent terms in the soft anomalous dimension 
$\gamma_{\cal S}$ for the Sudakov form factor, \eq{gamma_S_Sud_final}.
This is the contribution that needs to be combined with our jets in order to reconstruct the choice of Refs.~\cite{Sterman:2002qn,MertAybat:2006mz}, 
where jets are defined as square roots of the form factor.

\eq{ansatz_calS} can be compared with the direct computation of the soft anomalous dimension at one loop. For the latter we express the sum over all diagrams as
\beq 
\label{calS_direct_computation}
{\cal S}\left( 
\beta_i \cdot \beta_j, \as, \e \right) \, = \, 1 + \frac{\as}{4 \pi} \,
\sum_{i=1}^{n}\sum _{j,\,j \neq i} I_{i j}^{(1)} \left( \beta_i \cdot \beta_j, \e \right)  
\sum_a {\rm T}_i^{(a)} {\rm T}_j^{(a)} + {\cal O}(\alpha_s^2) \, ,
\eeq
where the basic one-loop integral, stripped of colour factors, is  
\beq
\label{basic_single_gluon_integral}
I_{ij}^{(1)} \left( \beta_i \cdot \beta_j, \e \right) \, = \, 
\frac{1}{\e^2} - \frac{1}{\e} \ln(\beta_i \cdot \beta_j \, 
{\rm e}^{{\rm i} \pi \lambda_{ij}} ) \, .
\eeq
The anomalous dimension corresponding to \eq{calS_direct_computation} is
\beq 
\label{summing_up_diagrams_1loop}
\Gamma^{\cal S} \left( 
\beta_i \cdot \beta_j, \as, \e \right) \, = \, 
\sum_{i=1}^{n}\sum _{j,\,j \neq i} \, \frac{\alpha_s}{2 \pi} \left[\frac{1}{\e} -
\ln (\beta_i \cdot \beta_j \, {\rm e}^{{\rm i} \pi \lambda_{ij}} )
\right] \, \sum_a {\rm T}_i^{(a)} {\rm T}_j^{(a)} + {\cal O}(\alpha_s^2) \, ,
\eeq
which coincides with \eq{ansatz_calS} upon substituting there the one-loop 
values of the anomalous dimensions, and upon using again colour conservation\footnote{Note that in explicit calculations in a given colour basis the fact that
\[
\left[ \sum_{j, \, j \neq i} \sum_a {\rm T}_i^{(a)} {\rm T}_j^{(a)}
\right]_{MN} = - C_i \delta_{MN} \, , 
\]
which is a statement of gauge invariance, provides a useful consistency 
check on the calculation of the colour factors.  A simple example is provided 
in \eq{qq_bar_1loop}: one can verify that the sum of the three matrices 
there yields $- C_F$ times the unit matrix, as it should.} for the 
$1/\epsilon$ term.

\section{Conclusions}
\label{conclu}

We have explored the singularity structure of fixed--angle scattering 
amplitudes in massless gauge theories using their factorization into 
gauge--invariant hard, soft and jet functions. The factorization formula we 
use, \eq{facamp}, is based exclusively on dimensional regularisation. It has 
the advantage that Wilson line correlators, such as the soft function 
${\cal S}$,  are pure counterterms to all orders in perturbation theory, and 
they do not depend on any mass scales.

Our central observation is the presence of a symmetry under rescaling of 
the velocities that define the eikonal functions ${\cal S}$ and ${\cal J}$, 
\eq{rescaling}; this symmetry is built into the eikonal Feynman rules, but 
it is broken for any eikonal function involving light-like segments, due to 
the cusp anomaly. Multi-parton scattering amplitudes depend on a specific
combination of eikonal soft and jet functions, the reduced soft function of 
\eq{reduS}, where the cusp anomaly cancels and the rescaling symmetry is 
restored. We have shown that the specific way in which this symmetry is 
broken and then restored imposes tight constraints on the functional 
dependence of these functions on the kinematic variables, as well as on 
the colour variables. The all-order constraints on the reduced soft function 
are summarized by \eq{oureq}.

For purely eikonal functions which do not depend on any kinematic scale and 
are affected by the cusp anomaly, we observe that the complete kinematic 
dependence of the single pole terms in the exponent is tightly connected to 
the double poles, and they are both associated with the cusp anomaly. This is 
easy to see for eikonal functions involving two rays, such as the eikonal jet or 
the eikonal version of the Sudakov form factor, but the result turns out to be 
more general, and through \eq{oureq} it carries over to the multi-leg case. This 
is contrasted with partonic amplitudes, such as the partonic jet function,
\eq{final_J}, that have more complicated dependence on the kinematic 
variables, since they depend on  dimensionful scales and have non-trivial
renormalization properties. 

Our conclusions concerning large--angle soft singularities based on \eq{oureq}
can be summarized as follows: for amplitudes involving two or three hard 
external partons (and any number of non-coloured particles) the kinematic
dependence of the soft function is completely determined, and it is governed, 
to all orders in perturbation theory, by the cusp anomalous dimension. We
emphasize that this conclusion does not depend on the way in which the cusp
anomalous dimension itself depends on the colour representation of a given 
parton. The results for the anomalous dimensions of the reduced soft matrices 
in these cases are given in \eq{gamsbarres} and \eq{solution_n3_implicit}.

For amplitudes involving four or more hard external partons the available
constraints on the soft function are insufficient to uniquely determine the 
kinematic dependence. They nevertheless relate it to the cusp anomalous
dimension. Considering the component of the cusp anomalous dimension that 
is proportional to the quadratic Casimir, we have found a minimal solution, 
\eq{ansatz}, for the reduced soft anomalous dimension or, equivalently,
\eq{ansatz_calS} for the original soft anomalous dimension. This solution, 
which is valid for any number of legs, satisfies the all-order constraints and 
is consistent with all explicit calculations available to date. \eq{ansatz} is 
written as a sum over colour dipoles: it correlates the kinematic dependence 
with the dependence on the colour variables for each pair of hard partons in 
the very same way as one-loop diagrams do. In contrast with the two- and 
three-leg cases, for four or more legs the solution is not unique: the soft 
anomalous dimension may receive additional contributions (at three loops 
or beyond) which however depend on kinematic variables exclusively through
conformal cross ratios. We point out that the absence of such contributions 
at two loops is non trivial: terms that satisfy this requirement, displayed in 
\eq{hat_H}, do appear in two-loop diagrams at ${\cal O}(1/\epsilon)$, but 
the calculation of Ref.~\cite{MertAybat:2006mz} proves that they do not
contribute to the soft anomalous dimension at this order. This suggests that 
\eq{ansatz} is in fact the full answer to any loop order, aside from corrections
(which in turn satisfy \eq{oureq_reformulated_HC}) that are induced by 
possible contributions of higher-order Casimir operators to the cusp 
anomalous dimension itself, which may appear at four loops or beyond.

To summarize, we took here a step towards the understanding of the infrared
singularity structure of gauge theory amplitudes to all orders in perturbation 
theory. We did so by identifying a rescaling symmetry of eikonal functions appearing in the factorization of the amplitudes, which is broken by the cusp
anomaly, and then restored in a specific way. Using this observation, we 
derived all-order constraints on soft anomalous dimension matrices, that are 
valid for any number of external partons. We then studied the consequences 
of these constraints and established the complete solution for the soft 
anomalous dimension for amplitudes involving two or three partons, and a 
minimal solution for four partons or more. By formulating our results for the 
jet and soft functions in terms of a few universal anomalous dimensions, which
depend only on the dimensionally--regularized coupling, and not on kinematics,
or on a colour basis, we significantly increase the predictive power of the
factorization formula, providing powerful checks on multi-loop calculations 
and a better starting point for soft gluon resummation.

\newpage

\noindent
{\bf Note added}

\noindent
A few hours before the submission of our paper to the arXiv, T. Becher and M. Neubert published an independent study~\cite{Becher:2009cu}, proposing
an ansatz for the soft anomalous dimension matrix which is essentially
equivalent to our \eq{ansatz_calS}.

\acknowledgments

We would like to thank Lance Dixon, George Sterman and Gregory Korchemsky for discussions. L.M. would like to thank the School of Physics of the University of Edinburgh for hospitality during the early stages of this work. Work supported in part by MIUR under contract 2006020509$\_$004,  and by the European Community's Marie-Curie Research Training Network `Tools and Precision Calculations for Physics Discoveries at Colliders'  (`HEPTOOLS'), under contract MRTN-CT-2006-035505.

\vspace{1cm}

\appendix

\section{The case of amplitudes with three partons}
\label{three_partons}

It is well known that the colour structure of the soft function for amplitudes 
with three hard partons is trivial: the soft matrix is proportional to the identity
matrix. Here we briefly explain this property and observe another special 
property of the soft function in this case: similarly to the $n = 2$ case,
discussed in \secn{conssud}, for $n = 3$ the contraints of \eq{oureq} 
completely determine the dependence on the kinematics.

The reduced soft matrix, defined in \eq{reduS}, in this case is given by
\beq
 \overline{{\cal S}}^{[f]}_{MN} \left(\rho_{12}, \, \rho_{23}, \, \rho_{31},
 \, \as, \e \right) = \frac{{\cal S}^{[f]}_{MN} \left(\beta_1 \cdot \beta_2,
 \, \beta_2 \cdot \beta_3, \, \beta_3 \cdot \beta_1, \, \as, \e 
 \right)}{ {\cal J}_{1} \left(\frac{2 (\beta_1 \cdot  n_1)^2}{n_1^2}, 
 \as, \e \right) \, {\cal J}_{2} \left(\frac{2 (\beta_2 \cdot n_2)^2}{n_2^2}, 
 \as, \e \right) \, {\cal J}_{3} \left(\frac{2 (\beta_3 \cdot n_3)^2}{n_3^2}, 
 \as, \e \right)} \, .
\label{reduS_n3}
\eeq
We now note that \eq{oureq} yields three independent differential equations, 
which are linear in $\ln (\rho_{ij})$. Explicitly, they are
\beqa
  \left[\frac{\partial}{\partial \ln(\rho_{12})} \, + \, 
         \frac{\partial}{\partial \ln(\rho_{31})}
  \right] 
  \Gamma^{{\overline{\cal S}}}_{MN} \left(\rho_{12}, \, \rho_{23}, \, \rho_{31}, 
  \as \right) 
  & = & \frac{1}{4} \delta_{MN}  \, {\gamma}_K^{(1)} 
  \left( \as \right) \, , \nonumber \\
\label{oureq_reformulated_n3}
  \left[\frac{\partial}{\partial \ln(\rho_{23})} \, + \, 
         \frac{\partial}{\partial \ln(\rho_{12})}
  \right] 
  \Gamma^{{\overline{\cal S}}}_{MN} \left(\rho_{12}, \, \rho_{23}, \, \rho_{31}, 
  \as \right) 
  & = & \frac{1}{4} \delta_{MN}  \, {\gamma}_K^{(2)} 
  \left( \as \right) \, , \\
  \left[\frac{\partial}{\partial \ln(\rho_{23})} \, + \, 
         \frac{\partial}{\partial \ln(\rho_{31})}
  \right] 
  \Gamma^{{\overline{\cal S}}}_{MN} \left(\rho_{12}, \, \rho_{23}, \, \rho_{31},
  \as \right) 
  & = & \frac{1}{4} \delta_{MN}  \, {\gamma}_K^{(3)}
  \left( \as \right) \, . \nonumber
\eeqa
Having three independent linear equations in the three variables there is a unique solution. It is given by
\begin{align}
\begin{split}
\label{solution_n3_implicit}
\Gamma^{\overline{S}}_{MN}
\left( \rho_{i j}, \as \right) & = \, \bigg\{- \frac18 \, 
\bigg[
\left({\gamma}_K^{(3)} - {\gamma}_K^{(1)} - {\gamma}_K^{(2)} \right) \ln(\rho_{12}) +
\left({\gamma}_K^{(1)} - {\gamma}_K^{(2)} - {\gamma}_K^{(3)} \right) \ln(\rho_{23}) \\ & +
\left({\gamma}_K^{(2)} - {\gamma}_K^{(3)} - {\gamma}_K^{(1)} \right) \ln(\rho_{13})
\bigg] \, + \, \frac12  \,
\left[\delta_{{\overline{\cal S}}}^{(1)}  + \delta_{{\overline{\cal S}}}^{(2)} 
+ \delta_{{\overline{\cal S}}}^{(3)} \right] \bigg\} \,\delta_{MN} \, ,
\end{split}
\end{align} 
where a constant term was added, as in \eq{ansatz}. Note that we have
shown that $\Gamma^{\overline{S}}_{MN}$ is proportional to the unit matrix in
colour space without specifying the representation of the partons. This result is completely general, and it generalizes the previously known two-loop result to 
all orders in perturbation theory. Just as in the case of $n = 2$, the entire
kinematic dependence is controlled by the cusp anomalous dimension. As 
explained in Sections \ref{const} and \ref{ansa}, the fact that uniqueness can 
be established for $n = 2, 3$ but not for $n \geq 4$ is related to the absence
of conformal cross ratios of the form of \eq{conformal_ratio_rho}.

In \eq{solution_n3_implicit}, the dependence on the representation of 
the various partons is implicit, appearing through the functions 
${\gamma}_K^{(i)}(\as)$ and $\delta_{{\overline{\cal S}}}^{(i)}(\as)$.
As discussed in~\secn{eikojet}, these may include higher--order corrections 
that are not proportional to the quadratic Casimir. In this respect the result 
of \eq{solution_n3_implicit} goes beyond the ansatz of \eq{ansatz}, which
considers only terms that are associated with the quadratic Casimir 
contributions to $\gamma_K^{(i)}$ in \eq{gamma_K}. Of course, ignoring 
$\widetilde{\gamma}_K$ in \eq{gamma_K} one recovers \eq{ansatz}.
Upon substituting \eq{ansatz} into \eq{oureq_reformulated_n3}, these 
three equations yield
\begin{align}
\label{n3_colour}
\begin{split}
  \sum_a {\rm T}_1^{(a)} \left({\rm T}_2^{(a)} + {\rm T}_3^{(a)}
  \right) & = - \, C_1 \, , \\
  \sum_a {\rm T}_2^{(a)} \left({\rm T}_3^{(a)} + {\rm T}_1^{(a)}
  \right) & = - \, C_2 \, , \\
  \sum_a {\rm T}_3^{(a)} \left({\rm T}_1^{(a)} + {\rm T}_2^{(a)}
  \right) & = - \, C_3 \, ,
\end{split}
\end{align}
which are satisfied owing to colour conservation, \eq{T_prop2}. 
\eq{n3_colour} also implies that
\begin{align}
\label{colour_factor_results_n3}
\begin{split}
2 \sum_a {\rm T}_1^{(a)} {\rm T}_2^{(a)} & = C_3 - C_1 - C_2 \, , \\
2 \sum_a {\rm T}_2^{(a)} {\rm T}_3^{(a)} & = C_1 - C_2 - C_3 \, , \\
2 \sum_a {\rm T}_3^{(a)} {\rm T}_1^{(a)} & = C_2 - C_1 - C_3 \, ,
\end{split}
\end{align}
which is consistent with the observation that all colour factors entering \eq{ansatz} in this case are proportional to the unit matrix. Finally, the explicit sum-over-dipoles solution to \eq{oureq_reformulated_n3} takes the form
\begin{align}
\begin{split}
\label{solution_n3_QC}
\Gamma^{\overline{S}}_{\text{Q.C.}}
\left( \rho_{i j}, \as \right) & = \, - \frac18 \, \widehat{\gamma}_K
\left( \as \right) \bigg[
\left(C_3 - C_1 - C_2 \right) \ln(\rho_{12}) +
\left(C_1 - C_2 - C_3 \right) \ln(\rho_{23}) \\ & +
\left(C_2 - C_3 - C_1 \right) \ln(\rho_{13})
\bigg] \, + \, \frac12 \, \widehat{\delta}_{{\overline{\cal S}}} (\as) \,
(C_1 + C_2 + C_3) \, .
\end{split}
\end{align}

\section{The case of $q\bar{q}\to q\bar{q}$ scattering at one loop}
\label{qqbar-one-loop-example}

The four-parton amplitude $q \bar{q} \to q \bar{q}$ provides a simple 
example where the colour matrix structure is non-trivial. We perform the 
calculation along the lines of Sec. IV of Ref.~\cite{MertAybat:2006mz}, 
but factorize the amplitude as in \eq{facamp}, using light-like Wilson lines.

The velocities are defined by:
\[
q (\beta_1) + \bar{q} (\beta_2) \to q (\beta_3) + \bar{q} (\beta_4) \, .
\]
It is convenient to set $\beta_1 = - v_1, \beta_2 = - v_2, \beta_3 = v_3$
and $\beta_4 = v_4$, so that all the scalar products $v_i \cdot v_j > 0$.
Note that below we will formally treat the four velocities as independent
variables, choosing not to enforce explicitly momentum conservation.
 
Following Ref.~\cite{MertAybat:2006mz}, we pick the colour basis 
\beq
\label{qq_basis}
c_1 = \delta_{12} \, \delta_{34} \, ;
\qquad \qquad c_2 = \delta_{13} \, \delta_{24} \, ,
\eeq
and we use the convention of \eq{softcorr} to write the result in a matrix 
form. There are six one-loop diagrams altogether, and for each one of them 
the loop integral yields \eq{basic_single_gluon_integral}. Computing the colour
factors in the chosen basis and summing up the contributions of the six 
diagrams according to \eq{calS_direct_computation} we get
\beqa
\label{qq_bar_1loop}
\nonumber
&& {\cal S} \left( v_i \cdot v_j, \as, \e \right) \, = \, 1 \, + \, 
\frac{\as}{2 \pi} \left\{ \left(
\begin{array}{cc}
\frac{1}{2 N_c} & 0 \\
- \frac12       & - C_F
\end{array} \right)
\left[ \frac{2}{\e^2} - \frac{1}{\e} \ln(v_1\cdot v_3) - 
\frac{1}{\e} \ln(v_2 \cdot v_4) \right] \right. \\ 
&& \hspace{1cm}  + \,\left(
\begin{array}{cc}
- \frac{1}{2 N_c} & \frac12 \\
\frac12       & - \frac{1}{2 N_c}
\end{array} \right)
\left[ \frac{2}{\e^2} - \frac{1}{\e} \ln(v_1 \cdot v_4) - 
\frac{1}{\e} \ln(v_2 \cdot v_3) \right] \\ 
&& \hspace{1cm} + \, \left. \left(
\begin{array}{cc}
- C_F & - \frac12 \\
0    & \frac{1}{2 N_c}
\end{array}\right)
\left[ \frac{2}{\e^2} - \frac{1}{\e} \left(\ln(v_1 \cdot v_2) + 
{\rm i} \pi \right) - \frac{1}{\e} \left(\ln(v_3 \cdot v_4) + {\rm i} \pi \right)
\right] \right\} \, + \, {\cal O}(\alpha_s^2) \, . \nonumber
\eeqa
The corresponding one-loop soft anomalous dimension matrix is then
\begin{align}
\label{Gamma_S_qqbar_scattering}
\begin{split}
& \Gamma^S (v_i \cdot v_j, \alpha_s, \e)
= \frac{\alpha_s}{\pi} \, C_F \, \left(
\begin{array}{cc}
- \frac{2}{\epsilon} + \ln \big( ( v_1 \cdot  v_2) \, ( v_3 \cdot v_4) \big)
& 0 \\ 0 & - \frac{2}{\epsilon} + \ln \big( ( v_1 \cdot  v_3) \, 
(v_2 \cdot  v_4) \big)
\end{array}
\right) \\ & + \, \frac{\alpha_s}{2\pi}
\left(
\begin{array}{cc}
\frac{1}{N_c} \ln \left(\frac{( v_1 \cdot v_4) \, ( v_2 \cdot  v_3)}{(v_1
\cdot  v_3) \, ( v_2 \cdot  v_4)} \right) 
&  \ln \left(\frac{( v_1 \cdot  v_2) \, ( v_3 \cdot v_4)}{( v_1 \cdot  v_4)
\, ( v_2 \cdot v_3)} \right) \\
\ln \left(\frac{( v_1 \cdot  v_3) \, ( v_2 \cdot v_4)}{( v_1 \cdot  v_4)
\, ( v_2 \cdot v_3)} \right)
& \frac{1}{N_c} \ln \left(\frac{( v_1 \cdot  v_4) \, ( v_2 \cdot v_3)}{(v_1
\cdot  v_2) \, ( v_3 \cdot v_4)} \right)
\end{array}
\right) \,
- { 2 \pi {\rm i}} \, \, \frac{\alpha_s}{\pi} \, \left(
\begin{array}{cc}
- C_F & - \frac12 \\
0    & \frac{1}{2 N_c}
\end{array}\right)
\, + \, {\cal O}(\alpha_s^2) \, .
\end{split}
\end{align}
Finite terms agree with Eq. (4.21) in Ref.~\cite{MertAybat:2006mz}. Note 
that in that paper there are no poles in $\Gamma^S$ since the regularization 
used takes the Wilson lines off the light cone.
 
In order to compute the anomalous dimension matrix for the reduced soft 
function, we should subtract the jet anomalous dimensions, according to 
\eq{tran},
\beq
\Gamma^{\overline{{\cal S}}}_{I J} \left(\rho_{i j}, \as \right)
= \Gamma^{{\cal S}}_{I J} \left( \beta_i \cdot \beta_j, \as, \e \right) - 
\delta_{I J} \sum_{k = 1}^{4} \gamma_{{\cal J}_{k}} \left(w_k, 
\as, \e \right) \, .
\eeq
The jet anomalous dimensions $\gamma_{{\cal J}_{k}}$ can be 
computed using \eq{fingamJ} in the fundamental representation. At 
one loop this yields
\beq
\gamma_{{\cal J}_{k}} \left(w_k, \as, \e \right)  = 
\frac{\as}{2 \pi} \, C_F \, \left[ - 1 
+ \, \ln \left(\frac{2 (v_k \cdot n_k)^2}{n_k^2} \right) \, 
- \, \frac{1}{\epsilon} \, \right]  \, + \, {\cal O}(\alpha_s^2) \, .
\eeq
The subtracted terms are proportional to the unit matrix in colour space, so 
they affect only the diagonal elements of the anomalous dimension matrix.
We end up with the following anomalous dimension for ${{\overline{\cal S}}}$, 
which is of course finite,
\beqa
\label{Gamma_barS_qqbar_scattering}
\Gamma^{{\overline{\cal S}}}(\rho_{ij},\alpha_s) & = & 
\frac{\alpha_s}{\pi} \, C_F \, \left(
\begin{array}{cc}
2 + \frac12 \ln \left(\rho_{12} \, \rho_{34} \right) & 0 \\
0 & 2 + \frac12 \ln \left(\rho_{13} \, \rho_{24} \right)
\end{array}
\right)
\\ \nonumber
& + & \frac{\alpha_s}{4\pi} \left(
\begin{array}{cc}
\frac{1}{N_c} \ln \left(\frac{\rho_{14} \, \rho_{23}}{\rho_{13} \, \rho_{24}}
\right) 
&  \ln \left(\frac{\rho_{12} \, \rho_{34}}{\rho_{14} \, \rho_{23}} \right) \\
\ln \left(\frac{\rho_{13} \, \rho_{24}}{\rho_{14} \, \rho_{23}} \right) 
& \frac{1}{N_c} \ln \left(\frac{\rho_{14} \, \rho_{23}}{\rho_{12} \, 
\rho_{34}}\right) 
\end{array}
\right) \, - \, { 2 \pi {\rm i}} \, \frac{\alpha_s}{\pi} \,
\left(
\begin{array}{cc}
- C_F & - \frac12 \\
0    & \frac{1}{2 N_c}
\end{array} \right) \, + \, {\cal O}(\alpha_s^2) \, .
 \eeqa
It is straightforward to check that the general expressions in \eq{ansatz_calS} 
and \eq{ansatz} indeed reduce to \eq{Gamma_barS_qqbar_scattering} and
\eq{Gamma_S_qqbar_scattering}, respectively, upon evaluating the colour 
factors in the chosen basis and substituting the one-loop values for
$\gamma_K$, $\widehat{\delta}_{{\overline{\cal S}}}$ and 
$\widehat{\delta}_{\cal S}$.

\end{document}